\pdfoutput=1

\documentclass[twocolumn]{aastex63}
\usepackage{epstopdf}
\usepackage{lineno}
\usepackage{xspace}
\usepackage{graphicx}
\usepackage{amssymb}
\usepackage{amsmath}
\usepackage{natbib}
\usepackage{float}
\usepackage{hyperref}

\shorttitle{Chandra Imaging of the Western Hotspot in Pictor\,A}
\shortauthors{Thimmappa et al.}

\begin{document}

\title{{\it Chandra} Imaging of the Western Hotspot in the Radio Galaxy Pictor\,A: \\ Image Deconvolution and Variability Analysis}

\correspondingauthor{R.~Thimmappa}
\email{rameshan@oa.uj.edu.pl}

\author{R.~Thimmappa}
\affiliation{Astronomical Observatory of the Jagiellonian University, ul. Orla 171, 30-244 Krak\'ow, Poland}

\author{\L .~Stawarz}
\affiliation{Astronomical Observatory of the Jagiellonian University, ul. Orla 171, 30-244 Krak\'ow, Poland}

\author{V.~Marchenko}
\affiliation{Astronomical Observatory of the Jagiellonian University, ul. Orla 171, 30-244 Krak\'ow, Poland}

\author{K.~Balasubramaniam}
\affiliation{Astronomical Observatory of the Jagiellonian University, ul. Orla 171, 30-244 Krak\'ow, Poland}

\author{C.~C.~Cheung}
\affiliation{Naval Research Laboratory, Space Science Division, Washington DC 20375, USA}

\author{A.~Siemiginowska}
\affiliation{Harvard Smithsonian Center for Astrophysics, 60 Garden Street, Cambridge, MA 02138, USA}

\begin{abstract}
Here we present an analysis of the X-ray morphology and flux variability of the particularly bright and extended Western hotspot in the nearest powerful (FR\,II-type) radio galaxy, Pictor\,A, based on data obtained with the {\it Chandra} X-ray Observatory. The hotspot marks the position where the relativistic jet, that originates in the active nucleus of the system, interacts with the intergalactic medium, at hundreds-of-kiloparsec distances from the host galaxy, forming a termination shock that converts jet bulk kinetic energy to internal energy of the plasma. The hotspot is bright in X-rays due to the synchrotron emission of electrons accelerated to ultra-relativistic energies at the shock front. In our analysis, we make use of several {\it Chandra} observations targeting the hotspot over the last decades with various exposures and off-axis angles. For each pointing, we study in detail the PSF, which allows us to perform the image deconvolution, and to resolve the hotspot structure. In particular, the brightest segment of the X-ray hotspot is observed to be extended in the direction perpendicular to the jet, forming a thin, $\sim 3$\,kpc-long, feature that we identify with the front of the reverse shock. The position of this feature agrees well with the position of the optical intensity peak of the hotspot, but is clearly off-set from the position of the radio intensity peak, located $\sim 1$\,kpc further downstream. In addition, we measure the net count rate on the deconvolved images, finding a gradual flux decrease by about $30\%$ over the 15-year timescale of the monitoring.
\end{abstract}


\section{Introduction} 
\label{sec:intro}

In classical radio galaxies, bright hotspots are formed when relativistic jets, emanating from Active Galactic Nuclei (AGN), terminate due to interactions with the intergalactic medium (IGM) at large distances (tens- and hundreds-of-kiloparsecs) from host galaxies. These hotspots mark, in particular, the position of the termination shocks, where jet bulk kinetic energy is converted to internal energy of the plasma transported by the jets, and next injected into the extended lobes \citep{Blandford74}. The interaction between a relativistic but ``light'' jet and a much denser ambient medium leads, in fact, to the formation of a double-shock structure, consisting of a non-relativistic forward shock propagating into the ambient medium, and a mildly-relativistic reverse shock propagating within the outflow \citep[e.g.,][]{Kino04}. The intense non-thermal emission of the hotspots imaged from radio up to X-ray frequencies, is widely believed to originate in the near downstream of the reverse shock, where efficient acceleration of the jet particles to ultra-relativistic energies and magnetic field amplification likely take place \citep[e.g.,][]{Meisenheimer89,Stawarz07,Araudo16}.

\begin{deluxetable*}{ccccccccc}[t!]
\tabletypesize{\scriptsize}
\tablecaption{Observational data and spectral fitting results. \label{tab:ObsID}}
\tablehead{\colhead{ObsID} & \colhead{Date} & \colhead{MJD} & \colhead{Exposure} & \colhead{$\theta$} & \colhead{$\Gamma$}  & \colhead{red.${\chi}^2$} & \colhead{$F_{\rm 0.5-7.0\,keV}$} & \colhead{Counts$^{\dagger}$}\\
\colhead{} & \colhead{}  & \colhead{}  & \colhead{[ksec]} & \colhead{[arcmin]} & \colhead{}  & \colhead{} & \colhead{[$10^{-13}$\,erg\,cm$^{-2}$\,s$^{-1}$]}  & \colhead{}}
\startdata
\\
		346 & 2000-01-18 & 51561 & 25.8 & 3.50 & $2.01 \pm 0.05$ &  0.272 & $5.41\,\, (+0.20\, /-0.45)$ & 3461\\
		3090 & 2002-09-17 & 52534 & 46.4 & 0.11 & $1.96 \pm 0.03$ & 0.377 & $5.64\,\, (+0.03\, /-0.20)$ & 5278\\
		4369 & 2002-09-22 & 52539 & 49.1 & 0.11 & $1.99 \pm 0.03$ & 0.426 & $5.61\,\, (+0.11\, /-0.15)$ & 5564\\
		12039& 2009-12-07& 55172 & 23.7 &3.35 & $1.98 \pm 0.06$ & 0.260& $5.71\,\, (+0.02\, /-0.10)$ & 2290\\
		12040& 2009-12-09 &55174 & 17.3 & 3.35 & $2.07 \pm 0.08$  & 0.265 &$5.47\,\, (+0.08\, /-0.25)$ & 1710\\
		11586& 2009-12-12 & 55177 & 14.3 & 3.35 & $2.11 \pm 0.09$ & 0.212 &$5.39\,\, (+0.21\, /-0.40)$ & 1427\\
		14357 &2012-06-17 & 56095 & 49.3 & 3.07 & $2.05 \pm 0.05$ & 0.321& $5.88\,\, (+0.06\, /-0.14)$ & 3043\\
		14221& 2012-11-06 & 56237 & 37.5 & 3.10 & $2.08 \pm 0.05$ & 0.356& $5.84\,\, (+0.01\, /-0.11)$ & 3248\\
		15580& 2012-11-08& 56239 & 10.5& 3.10 & $2.08 \pm 0.14$ & 0.235& $5.24\,\, (+0.30\, /-0.21)$ & 935\\
		14222& 2014-01-17& 56674 & 45.4 & 3.30 & $2.00 \pm 0.05$ & 0.329& $5.78\,\, (+0.12\, /-0.10)$ & 3428\\
		16478& 2015-01-09& 57031 & 26.8& 3.32 & $1.95 \pm 0.08$ &0.232& $5.30\,\, (+0.15\, /-0.54)$ & 1657\\
		17574& 2015-01-10 &57032 & 18.6 & 3.32 & $2.04 \pm 0.11$  &0.209& $5.33\,\, (+0.26\, /-0.21)$ & 1187 \\
		\\
\enddata
\tablecomments{$^{\dagger}$ total number of counts within the $0.5-7.0$\,keV range from a circular region with a radius 20\,px centered on the hotspot.}
\end{deluxetable*}

In the majority of cases, relatively small angular sizes and fluxes of the hotspots, in cosmologically distant radio galaxies, preclude any detailed morphological analysis of the termination shocks, with the available instruments characterized by a limited sensitivity and arc-second angular resolution. For such, one can only apply a detailed spectral analysis and modeling of the broad-band spectral energy distribution, utilizing the integrated radio, infrared, optical, and X-ray flux measurements \citep[e.g.,][]{Meisenheimer97,Brunetti03,Georganopoulos03,Hardcastle04,Tavecchio05,Werner12}. Only in a few cases of the brightest and most extended hotspots, sub-arcsec imaging at high-frequency radio and near-infrared/optical frequencies have been performed; besides the famous Cygnus\,A example \citep[see, e.g.,][]{Carilli96,Pyrzas15,Dabbech18}, these cases include 3C\,445 \citep{Prieto02,Orienti12,Orienti17}, and 4C\,+74.26 \citep{Erlund10}.

The other particularly bright and extended hotspot can be found at the edge of the Western lobe of the nearest FR\,II type \citep{Fanaroff74} radio galaxy, Pictor\,A, located at redshift $z=0.035$ \citep{Eracleous04}. The Western hotspot, which is much brighter than the Eastern hotspot located at the counter-jet side,  and which is located at the projected distance of about 4 arcmin from the nucleus, has been investigated in detail: at cm radio wavelengths using the Very Large Array \citep{Perley97}, at mid-infrared frequencies with the {\it Spitzer} Space Telescope \citep{Werner12} and the Wide-field Infrared Survey Explorer \citep[WISE;][]{Isobe17}, in near-infrared using ground-based data \citep{Meisenheimer97}, at optical wavelengths with the Faint Object Camera onboard the {\it Hubble} Space Telescope \citep{Thomson95}, and finally in X-rays with the {\it Chandra} X-ray Observatory \citep[][and references therein]{H16}. The hotspot was also the subject of high-resolution radio imaging with the Very Long Baseline Array \citep[VLBA;][]{Tingay08}.

In this paper, we reanalyze all available {\it Chandra} data for the Western hotspots in Pictor\,A. The data consist of 12 pointings, spread over 15 years starting from 2000, each characterized by a different exposure and a different off-axis angle. Our study is complementary to the analysis presented in \citet{H16}, in that we perform a detailed image deconvolution of the hotspot for each pointing in order to clearly resolve the hotspot structure and to investigate its flux variability, first reported by \citeauthor{H16}. 

Throughout the paper we assume the $\Lambda$CDM cosmology with $H_{0} = 70$\,km\,s$^{-1}$, $\Omega_{\rm m} = 0.3$, and $\Omega_{\Lambda} = 0.7$, for which the luminosity distance to the source is 154\,Mpc, and the conversion angular scale is 0.697\,kpc\,arcsec$^{-1}$. The photon index $\Gamma$ is defined as $F_{\varepsilon} \propto \varepsilon^{-\Gamma}$ for the photon flux spectral density $F_{\varepsilon}$ and the photon energy $\varepsilon$.

\begin{figure*}[!t]
\centering 
\includegraphics[width=\textwidth]{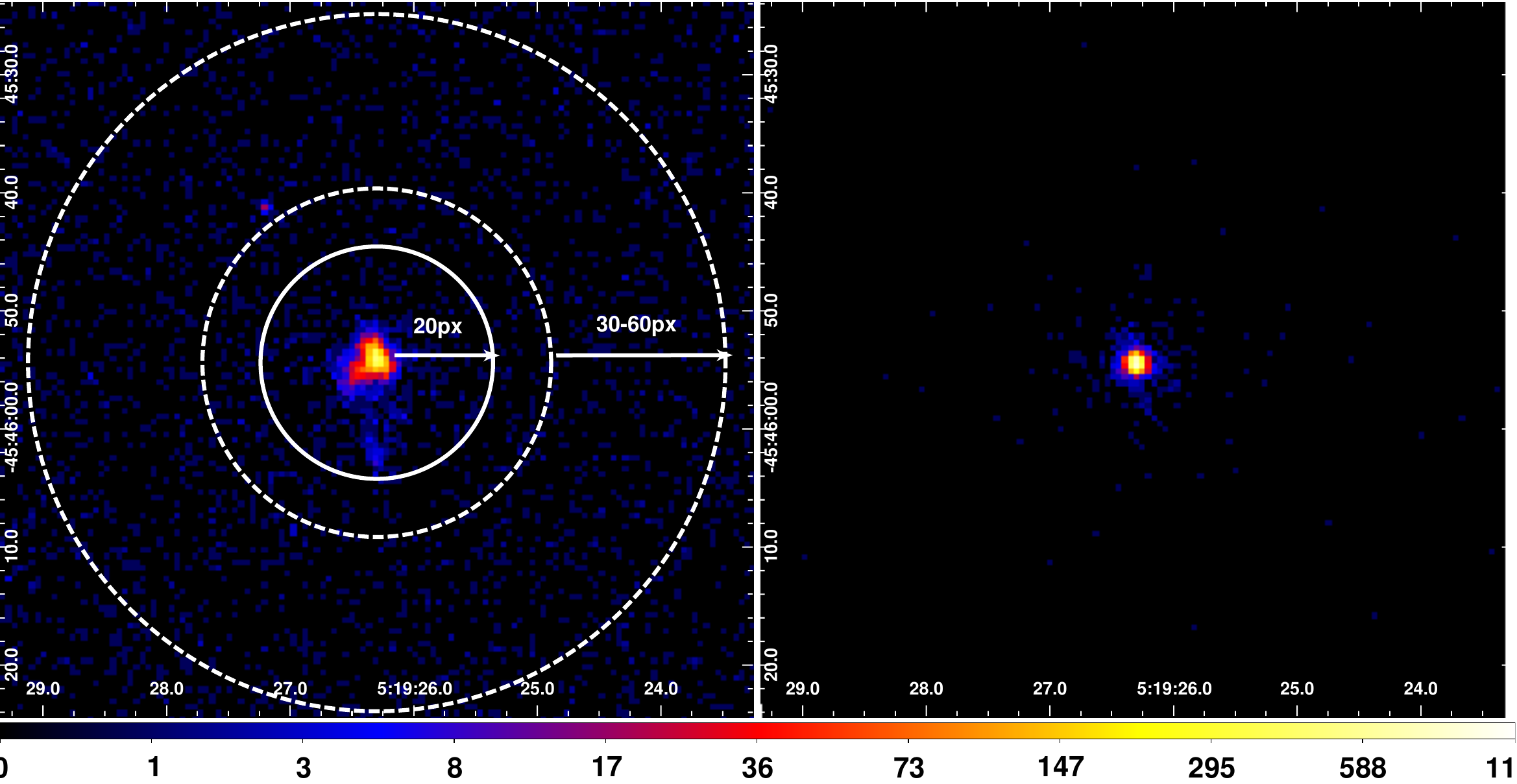}
\caption{{\it Left panel:} ACIS-S image of the W hotspot of Pictor\,A radio galaxy, within the energy range $0.5-7$\,keV for the ObsID\,3090 (with 1\,px binning). The source extraction region for the spectral modeling is denoted by the solid circle (20\,px radius), and the background annular region by dashed circles ($30-60$\,px). {\it Right panel:} the PSF simulated for the ObsID\,3090 (as discussed in Section\,\ref{sec:PSF}).}
\label{fig:source} 
\end{figure*}

\section{Chandra Data} 
\label{sec:data}

The Pictor\,A system has been observed with the Advanced CCD Imaging Spectrometer (ACIS), onboard {\it Chandra}, on 14 separate occasions over the past 15 years. The quality of the data for given regions of the source varies between the different pointings, due to the fact that the effective point spread function (PSF) is a complicated function of position in the imager, the source spectrum, and the exposure time. For our study we selected a 12-pointing subset of the \citet{H16} study, with a total observing time of 364\,ks; we excluded the two pointings (ObsID 14223 and 15593) for which the hotspot was located very close to the chip gap on the detector. In Table\,\ref{tab:ObsID} we list the ObsIDs, the dates and MJD of the observations, the exposure times, and the off-axis angles for the hotspot $\theta$. Note that the best-quality data for the hotspot corresponds to the ObsID 3090 and 4369, due to a combination of relatively long uninterrupted exposures (exceeding 45\,ksec) and small off-axis angles ($0.^{\prime}11 \simeq 7^{\prime\prime}$).

All data was reprocessed in a standard way with {\fontfamily{qcr} \selectfont CIAO}\,4.10 \citep{Fruscione06}, using the {\ttfamily chandra\_repro} script and the Calibration database CALDB\,4.7.8 recommended by the {\fontfamily{qcr} \selectfont CIAO} analysis threads. Pixel randomization was removed during the reprocessing and readout streaks were removed for all observational data. Point sources in the field were detected with the {\ttfamily wavdetect} tool and then removed for all analyzed ObsIDs. Throughout this paper, we selected photons in the $0.5-7.0$\,keV range. Counts and spectra were extracted for the source and the background regions from individual event files using the {\ttfamily specextract} script. Spectral fitting was done with the {\fontfamily{qcr}\selectfont Sherpa} package \citep{Freeman01}.

The total number of counts obtained for the W hotspot ($>30,000$) implies that calibration uncertainties dominate over statistical ones (see in this context \citealt{Drake06},  \citealt{Lee11},  \citealt{Lee11}, and the related discussion in \citealt{H16}). We note that given the observed X-ray flux of the source, there is also a possibility for a pile-up in the detector \citep{Davis01}, in particular in ObsID 3090 and 4369, for which the W hotspot was located at the center of the S3 chip, and the resulting count rates were $\simeq ˆ0.2$\,s$^{-1}$. Therefore, when performing the {\fontfamily{qcr}\selectfont MARX} simulation for the PSF modeling, we have accounted for the pile-up effect as well. 

\section{Data Analysis} 
\label{sec:analysis}

\subsection{Spectral modeling}
\label{sec:spectrum}

\begin{figure}[t!]
\centering 
\includegraphics[width=\columnwidth]{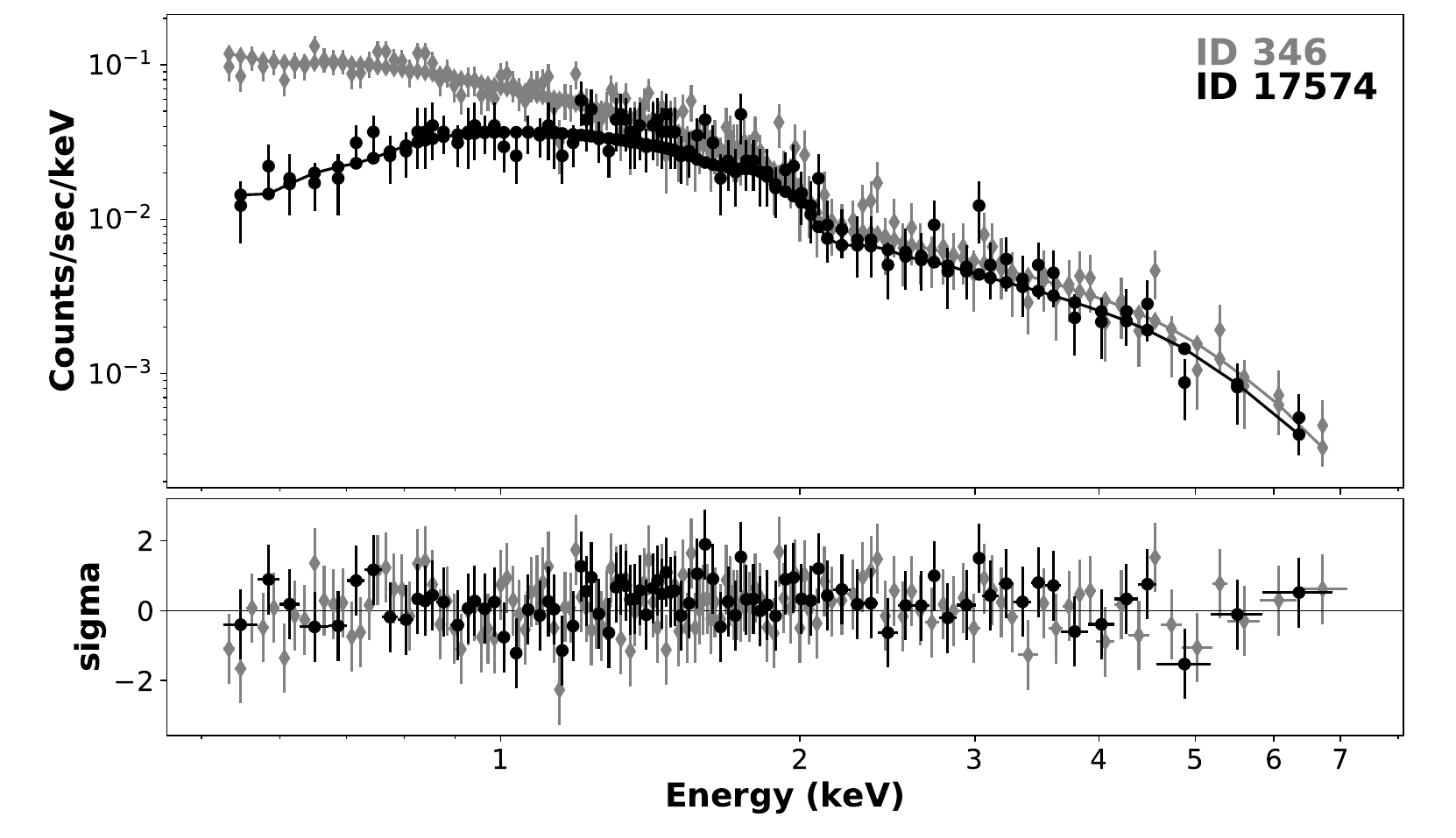}
\caption{The $0.5-7.0$\,keV spectra of the W hotspot of Pictor A, for ID 346 (gray diamonds) and 17574 (black circles), both fitted with the single power-law model moderated by the Galactic column density. Parameters of the model are given in Table\,\ref{tab:ObsID}.}
\label{fig:spectrum} 
\end{figure}

\begin{figure}[!t]
\centering 
\includegraphics[width=\columnwidth]{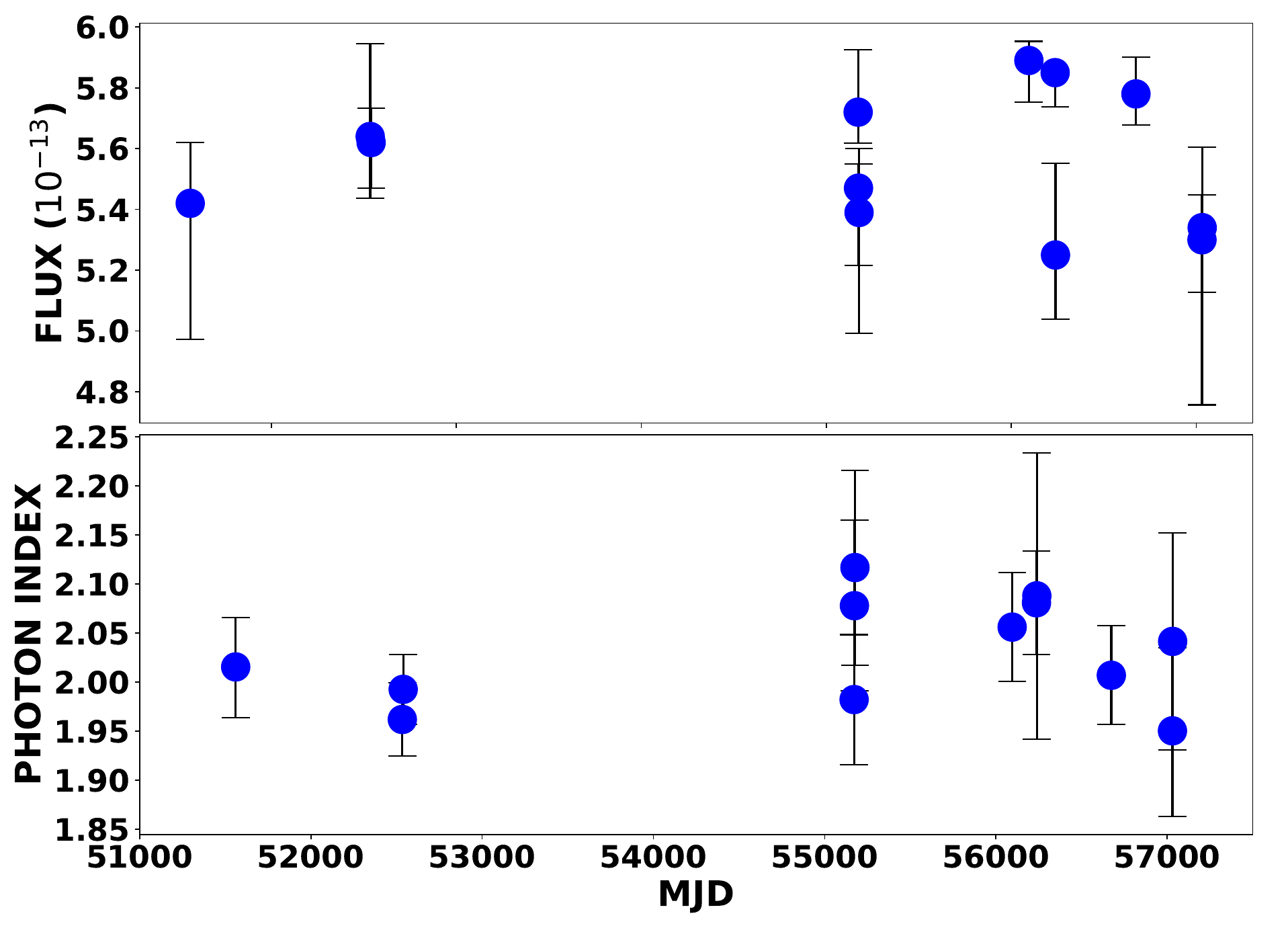}
\caption{The $0.5-7.0$\,keV energy flux (upper panel) and the photon index (lower panel) for the W hotspot of Pictor\,A, determined from single absorbed power-law model (see Table\,\ref{tab:ObsID}). The energy flux is given in the units of $10^{-13}$\,erg\,cm$^{-2}$\,s$^{-1}$.}
\label{fig:fits}
\end{figure}

For the spectral analysis, we have extracted the source spectrum in each ObsID from a circular region with a radius 20\,px ($\simeq 10^{\prime\prime}$, for the conversion scale $0.492^{\prime\prime}/{\rm px}$, and extracted the background spectrum from an annulus of $30-60$\,px ($\sim 15^{\prime\prime}-30^{\prime\prime}$), as illustrated in Figure\,\ref{fig:source} (left panel) for the ObsID\,3090. The background-subtracted source spectra (with a total number of net counts ranging from about 900 for the ObsID\,15580 up to about 5,000 for the ObsID\,4369, see Table\,\ref{tab:ObsID}) were fitted within the $0.5-7.0$\,keV range, assuming a single power-law model moderated by the Galactic column density $N_{\rm H,\,Gal} = 4.12 \times 10^{20}$\,cm$^{-2}$  \citep{Kalberla05}. Two examples of the fit, for the first exposure ObsID\,346 and the last exposure ObsID\,17574, are presented in Figure\,\ref{fig:spectrum}. The apparent difference between the low-energy segments of the spectra for the two pointings reflects, at least to some extent, the effective area of the detector at low photon energies decreasing with time \citep{Plucinsky18}. The results of the spectral fitting performed, for all the analyzed pointings, are summarized in Table\,\ref{tab:ObsID}, and visualized in Figure\,\ref{fig:fits}. 

Overall, in our spectral analysis we see that the single power-law model provides a reasonable description of the source spectrum, and is sufficient for single pointings when being fitted separately. In addition, even though we do see some changes in the $0.5-7.0$\,keV flux of the target between successive pointings, the source variability cannot be claimed at a high significance level, given the uncertainty in the flux estimates \citep[see in this context the discussion in][section 3.3 and Figure\,8 therein]{H16}. To highlight this point in quantitative, simple way, for our successive flux estimates $F_i \pm \sigma_i$ with $i=1,...,12$ (Table\,\ref{tab:ObsID}), we test for a constant signal by calculating
\begin{equation}
\chi^2 = \sum_{i=1}^{N=12} \frac{\left(F_i - F_{\rm m}\right)^2}{\sigma_i^2} \simeq 20.8 \, .
\end{equation}
Here the model $F_{\rm m}$ is the mean flux given the Gaussian errors (for which we assumed average values in the case of asymmetric errors reported in Table\,\ref{tab:ObsID}),
\begin{equation}
F_{\rm m} = \frac{\sum_{i=1}^{N=12} F_i/\sigma_i^2}{\sum_{i=1}^{N=12} 1/\sigma_i^2} \simeq 5.72 
\end{equation}
in units of $10^{-13}$\,erg\,cm$^{-2}$\,s$^{-1}$. The probability that our model is correct, is therefore
\begin{equation}
P\!({\chi^2}) = \frac{({\chi^2})^{(\nu-2)/2} \, \exp[-\chi^2/2]}{2^{\nu/2}\, \mathbf{\Gamma}\!(\nu/2)} \simeq 0.011 \, ,
\end{equation}
where $ \mathbf{\Gamma}\!(\nu/2)$ is the Gamma function, and the degrees of freedom $\nu = N-1 = 11$. As follows, $\chi^2/\nu \simeq 1.89$, and the corresponding probability for a constant flux reads as $\sim 1\%$. 

\subsection{PSF modeling}
\label{sec:PSF}

For modeling of the {\it Chandra} PSF at the position of the W hotspot of Pictor\,A, we have used the Chandra Ray Tracer ({\fontfamily{qcr} \selectfont ChaRT}) online tool \citep{Carter03}\footnote{\url{http://cxc.harvard.edu/ciao/PSFs/chart2/runchart.html}} and {\fontfamily{qcr} \selectfont MARX} software \citep{Davis12} \footnote{\url{https://space.mit.edu/cxc/marx}}. The centroid coordinates of the selected source region were taken as the position of a point source. The source spectrum for the spectral specification in {\fontfamily{qcr}\selectfont ChaRT} was the background-subtracted $0.5-7.0$\,keV spectrum created separately for each observation and fitted with a single power-law model, as described in the previous sub-section. Next, a collection of event files were made using {\fontfamily{qcr}\selectfont ChaRT} by tracing rays through the Chandra X-ray optics. Then the rays were projected onto the detector through {\fontfamily{qcr}\selectfont MARX} simulation, taking into account all relevant detector issues. Through this process, an event file was obtained from which an image of the PSF was created. We used one of the latest features in the last release of {\fontfamily{qcr}\selectfont MARX}, namely the option to use the energy-dependent sub-pixel event repositioning algorithm (EDSER) to adjust chip coordinates, and also to include the pileup effect. We reprocessed all the archival data, and used the aspect solution for each observation.

\begin{figure}[!t]
\centering 
\includegraphics[width=\columnwidth]{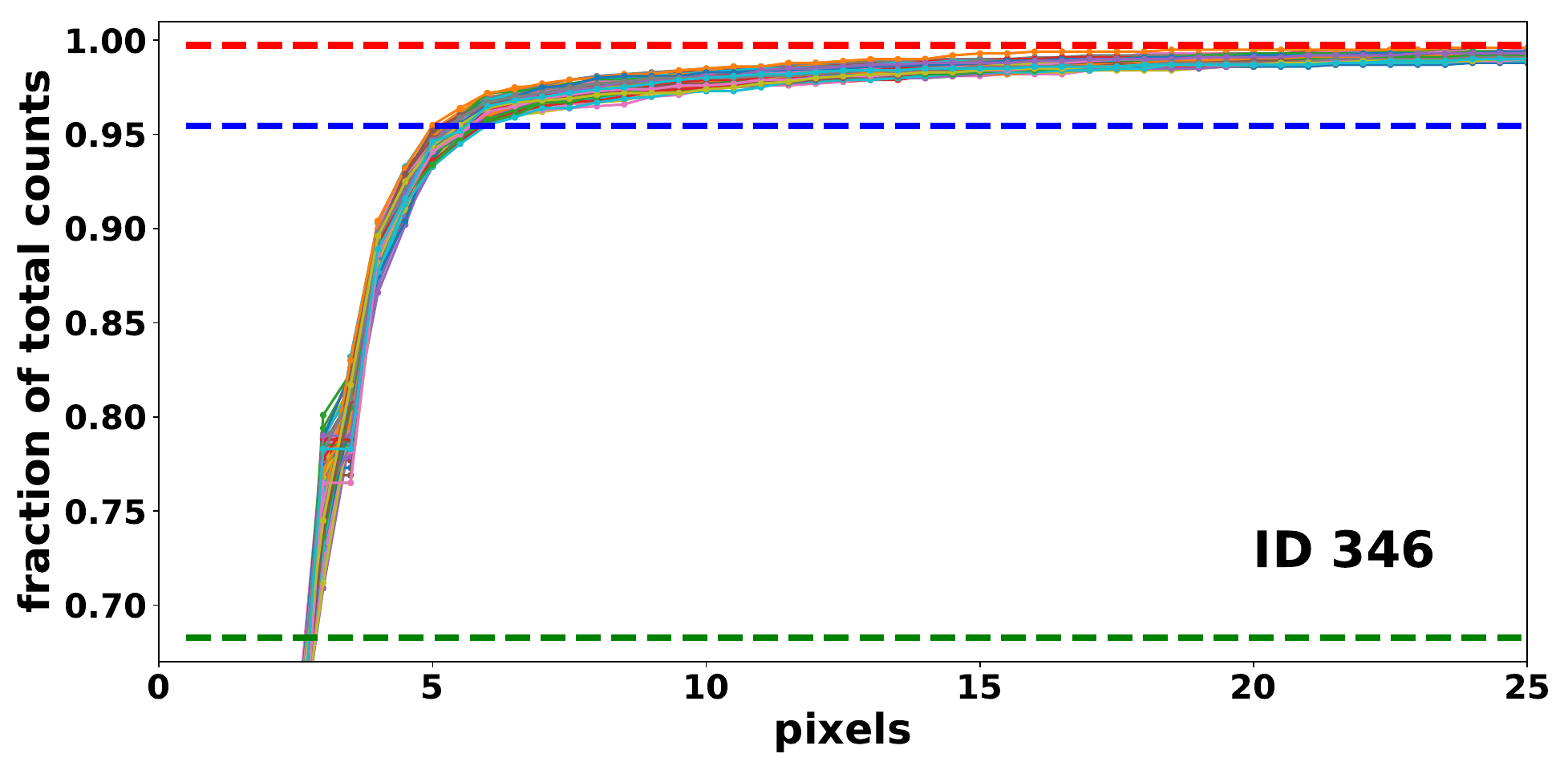}
\includegraphics[width=\columnwidth]{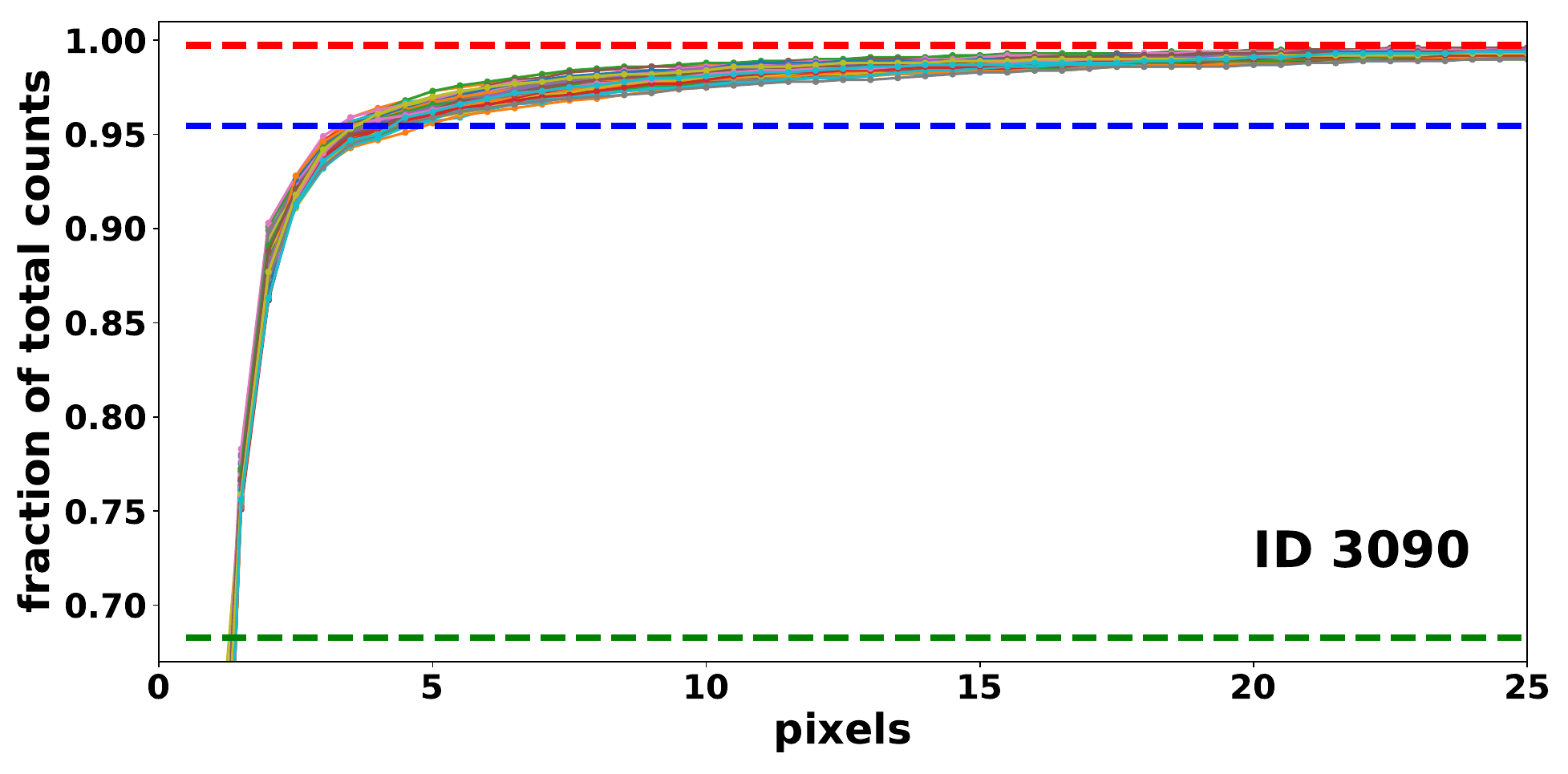}
\caption{The enclosed count fraction as a function of the radius aperture for the simulated PSFs for ObsIDs 346 (upper panel) and 3090 (lower panel). For each ObsID we performed 100 PSF simulations, and each curve corresponds to one particular PSF realization. The horizontal green, blue, and red lines (from bottom to top), correspond to $1\sigma$, $2\sigma$, and $3\sigma$ count fractions, respectively.}
\label{fig:ECF} 
\end{figure}

\begin{figure*}[!t]
\centering 
\includegraphics[width=\textwidth]{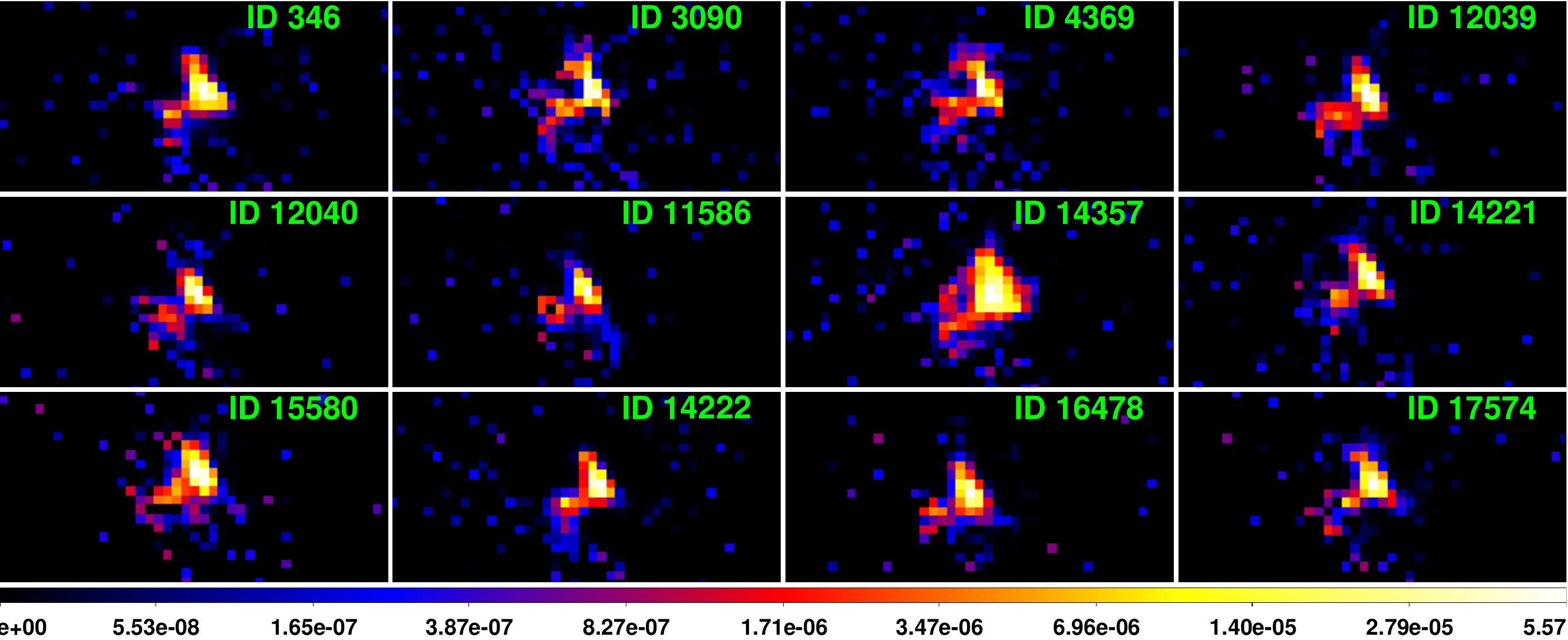}
\caption{The deconvolved {\it Chandra} images of the W hotspot in Pictor\,A at 1\,px resolution. Each image results from averaging over the restored images for 100 PSF realizations using the LRDA on the exposure-corrected maps. The color scale gives the count rate (cts\,s$^{-1}$).}
\label{fig:images-px}
\end{figure*}

\begin{figure*}[!t]
\centering 
\includegraphics[width=\textwidth]{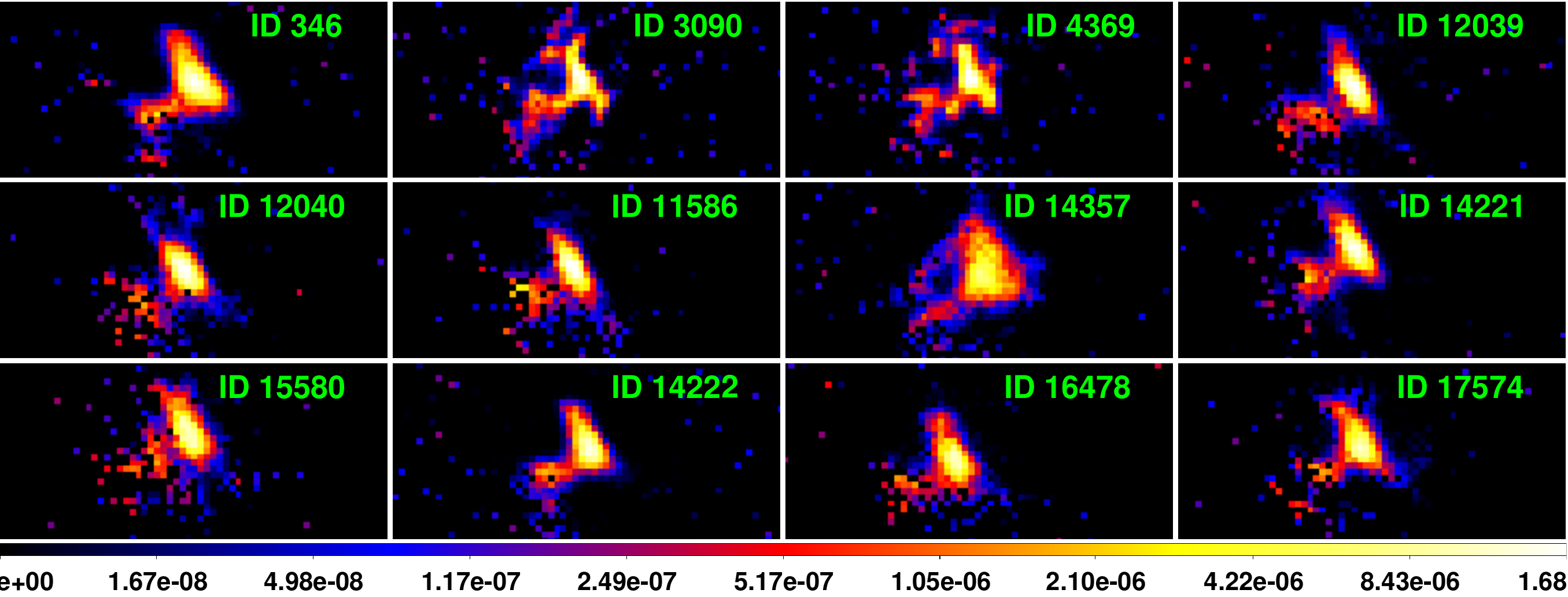}
\caption{The deconvolved {\it Chandra} images of the W hotspot in Pictor\,A at 0.5\,px resolution. Each image results from averaging over the restored images for 100 PSF realizations using the LRDA on the exposure-corrected maps. The color scale gives the count rate (cts\,s$^{-1}$).}
\label{fig:images-subpx}
\end{figure*}

\begin{figure*}[!t]
\centering 
\includegraphics[width=\textwidth]{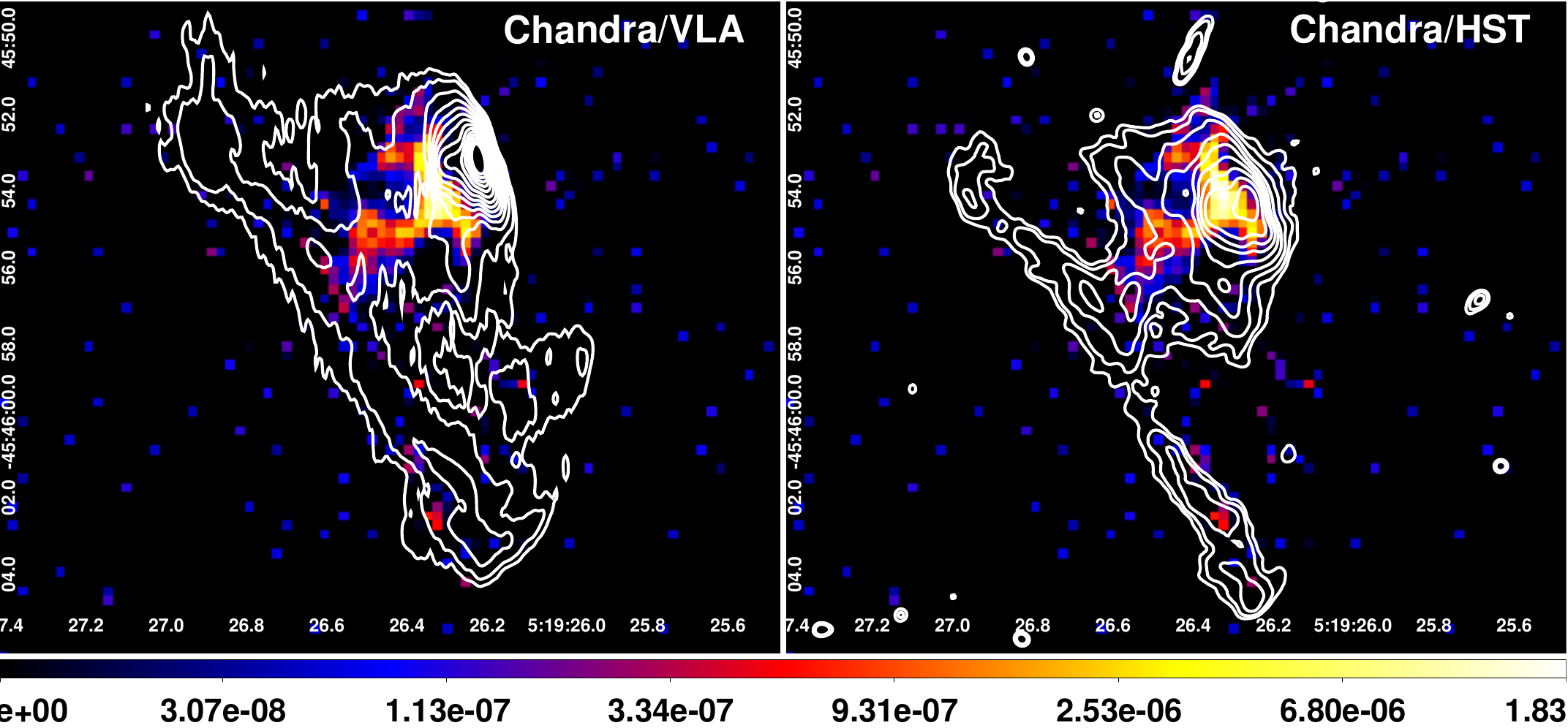}
\caption{The deconvolved exposure-corrected {\it Chandra} image of the W hotspot in Pictor\,A at 0.5\,px resolution for the ObsID 3090, averaged over 100 random realizations of the PSF, with the radio (3.6\,cm wavelength, beam size $0.^{\prime\prime}77 \times 0.^{\prime\prime}17$, position angle $-0.4$\,deg) VLA contours superimposed (left panel) and optical F606W filter ($5918\textup{\AA}$, $90\%$ encircled energy within radius $0.35^{\prime\prime}$) {\it Hubble} Space Telescope ACS/WFC contours superimposed (right panel). Radio contours are spaced with a factor of $\sqrt{2}$ between $0.552$ and $70.71\%$ of the peak intensity of $215$\,mJy\,beam$^{-1}$. Optical contours are spaced with a factor of $\sqrt{2}$ between $0.008$ and $3$\,cts\,s$^{-1}$.} 
\label{fig:radio}
\end{figure*}

For every ObsID, we perform the PSF simulations, as described above, 100 times, as each particular realization of the PSF is different due to random photon fluctuations; an example of the simulated PSF for the ObsID\,3090 is presented in Figure\,\ref{fig:source} (right panel). With such, we first study the enclosed count fraction (ECF) in the simulated PSFs, i.e. the fraction of counts that have been detected within a circular radius aperture (noting that the {\it Chandra}'s PSF is in reality more and more elongated with increasing off-axis angle). For illustration, in Figure\,\ref{fig:ECF} we present the resulting ECF as a function of the radius aperture for the first two ObsIDs 346 and 3090, which differ in both exposure time and off-axis angle (see Table\,\ref{tab:ObsID}). As shown in the figure, while in the case of ObsID 346 the $2\sigma$ radius is at $\gtrsim 5$\,px\,$\simeq 2^{\prime\prime}.5$, it is only $\lesssim 4$\,px\,$\simeq 2^{\prime\prime}.0$ for ObsID 3090; the $3\sigma$ radius extends in both cases beyond 25\,px\,$\simeq 12.^{\prime\prime}3$.

\subsection{Image Deconvolution}
\label{sec:deconvolve}

We use the Lucy-Richardson Deconvolution Algorithm (LRDA), which is implemented in the {\fontfamily{qcr}\selectfont CIAO} tool {\fontfamily{qcr}\selectfont arestore}, to remove the effect of the PSF and restore the intrinsic surface brightness distribution of the hot spot. The algorithm requires an image form of the PSF, which is provided by our {\fontfamily{qcr}\selectfont ChaRT} and {\fontfamily{qcr}\selectfont MARX} simulations as described in the previous sub-section, and exposure-corrected maps of the source \citep[see also in this context][for the {\it Chandra} image analysis of the large-scale jet in radio quasar 3C\,273]{Marchenko17}. As there are 100 iterations of PSF simulations for each given ObsID, we first produce 100 deconvolved images for each pointing, and next, average them. The collection of the resulting deconvolved images with 1\,px resolution are presented in Figure\,\ref{fig:images-px}, and with 0.5\,px resolution in Figure\,\ref{fig:images-subpx}.

We note that the deconvolved images corresponding to the ObsID\,14357 look very different than the deconvolved images for the other ObsIDs, being in particular blurred and seemingly ``unfocused''. Indeed, in this particular observation the hotspot was located on the S2 chip, while in the remaining observations analyzed here it was placed on the back illuminated S3 chip.

\begin{figure*}[!t]
\centering 
\includegraphics[width=\textwidth]{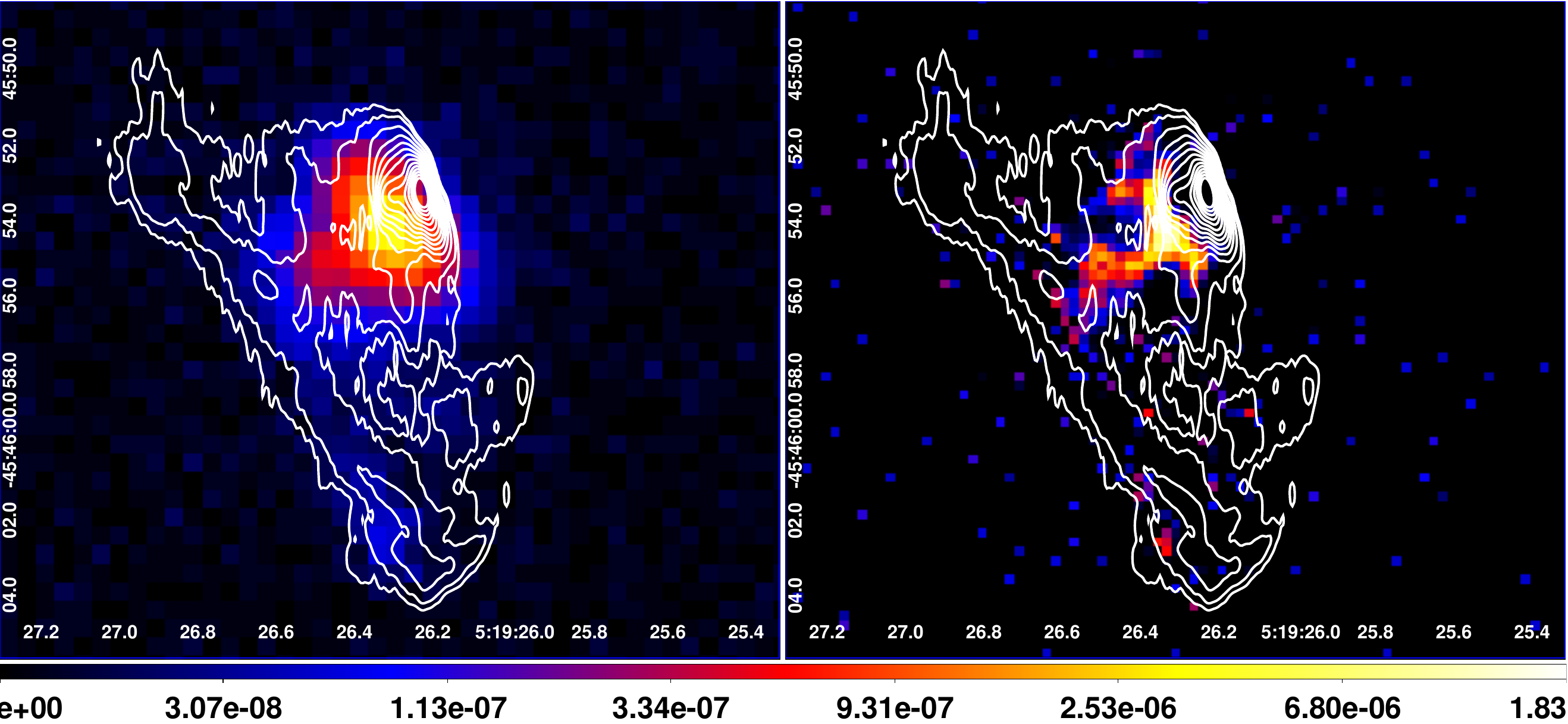}
\caption{A comparison between the {\it Chandra} image of the W hotspot in Pictor\,A resulting from merging all the analyzed ObsIDs excluding ObsID\,14357 (left panel), and the deconvolved exposure-corrected image at 0.5\,px resolution for the ObsID 3090, averaged over 100 random realizations of the PSF (right panel). In both panels, the VLA 3.6\,cm radio contours are superimposed (see Figure\,\ref{fig:radio}).} 
\label{fig:comparison}
\end{figure*}

\section{Discussion: Results of the Analysis} 
\label{sec:results}

In Figure\,\ref{fig:radio} (upper panel), we present the deconvolved exposure-corrected {\it Chandra} image of the W hotspot in Pictor\,A at 0.5\,px resolution for the ObsID 3090, averaged over 100 random realizations of the PSF, with the superimposed Very Large Array (VLA) radio contours \citep{Perley97}, as well as {\it Hubble} Space Telescope optical contours from archival ACS/WFC F606W data obtained in May 2015 (1.35\,hr exposure; program 13731). As shown, on the deconvolved {\it Chandra} image the X-ray structure of the hotspot is resolved into a seemingly linear feature oriented along the jet axis, and terminating into the perpendicular ($\sim 4^{\prime\prime}$-long) disk-like segment located just upstream ($\sim 1^{\prime\prime}.5$ away) of the intensity peak of the radio hotspot. We identify this segment with the Mach disk of the hotspot as discussed in \citet{Meisenheimer89}, i.e. the very front of the reverse shock formed within the jet plasma due to the interaction of the jet head with the ambient medium; alternatively, we could be looking at the upstream conical shock formed at the head of the jet, as seen in some runs of the hydrodynamical simulations of a light, supersonic outflow by \citet{Saxton02}, depending on the choice of the main model parameters (jet/ambient medium density contrast, jet Mach number, and jet viewing angle). The radio continuum emission peaks further downstream of this shock, and extends up the distance which, again, \emph{could} be dentify with the contact discontinuity where the shocked jet plasma meets the shocked IGM. The position of the optical intensity peak of the hotspot agrees well with the position of the X-ray disk. On the other hand, the filament --- a ``bar'' --- perpendicular to the jet axis, but located $\sim 10^{\prime\prime}$ upstream of the Mach disk, which is present in the radio image and particularly prominent in the optical image, possesses only a very weak X-ray counterpart on the deconvolved X-ray images. 

In Figure\,\ref{fig:comparison}, we show the X-ray image of the hotspot resulting from merging all the analyzed {\it Chandra} ObsIDs excluding ObsID\,14357 (left panel), and the deconvolved exposure-corrected image at 0.5\,px resolution for the ObsID 3090 (right panel). As shown, merging all the selected ObsIDs (aligned using the measured peak position of the hotspot) one significantly improves the photon statistics, but at the same time blurs the image due to different shapes of the PSFs superimposed. The offset between the X-ray and radio intensity peaks can, however, still be observed on the merged image \citep[see][]{H16}, although the X-ray structure of the hotspot is no longer resolved into the jet and the perpendicular Mach disk/conical shock.

\begin{figure}[!t]
\centering 
\includegraphics[width=\columnwidth]{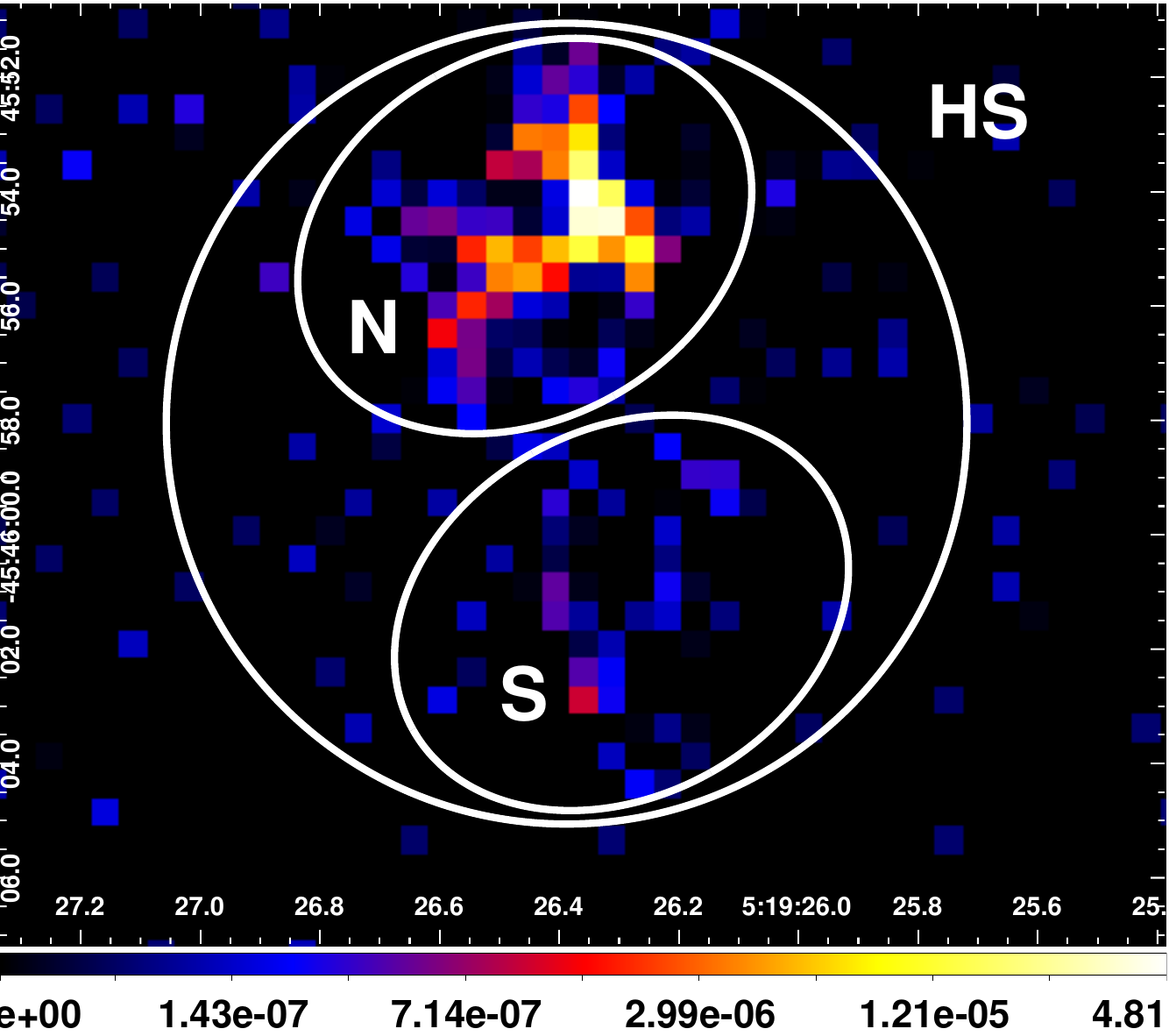}
\caption{The deconvolved exposure-corrected {\it Chandra} image of the W hotspot in Pictor\,A at 1.0\,px resolution for the ObsID 3090, averaged over 100 random realizations of the PSF. Source regions used for the extraction of the next counts --- ``hotspot total'' (HS), ``hotspot North'' (N), and ``hotspot South'' (S) --- are denoted by the circle and the two smaller ellipses.} 
\label{fig:regions}
\end{figure}

As the image deconvolution procedure should not affect the number of counts on the image, but only their distribution, we have additionally attempted to investigate the variability of the W hotspot by means of measuring the count rates on the obtained deconvolved (1\,px resolution) exposure-corrected maps of the target. In Figure\,\ref{fig:regions} we show the flux extraction regions defined for this purpose, including the circular ``hotspot total'' (HS) region with the 14\,px radius, the elliptical ``hotspot North'' (N) region with the major and minor axes of 8.5\,px and 6.5\,px, enclosing the main part of the hotspot, and finally the analogous elliptical ``hotspot South'' (S) region encompassing the Southern extension of the perpendicular filament (bar), which is pronounced clearly on the radio and optical maps, but only marginally on the {\it Chandra} images. 

\begin{figure}[!t]
\centering 
\includegraphics[width=\columnwidth]{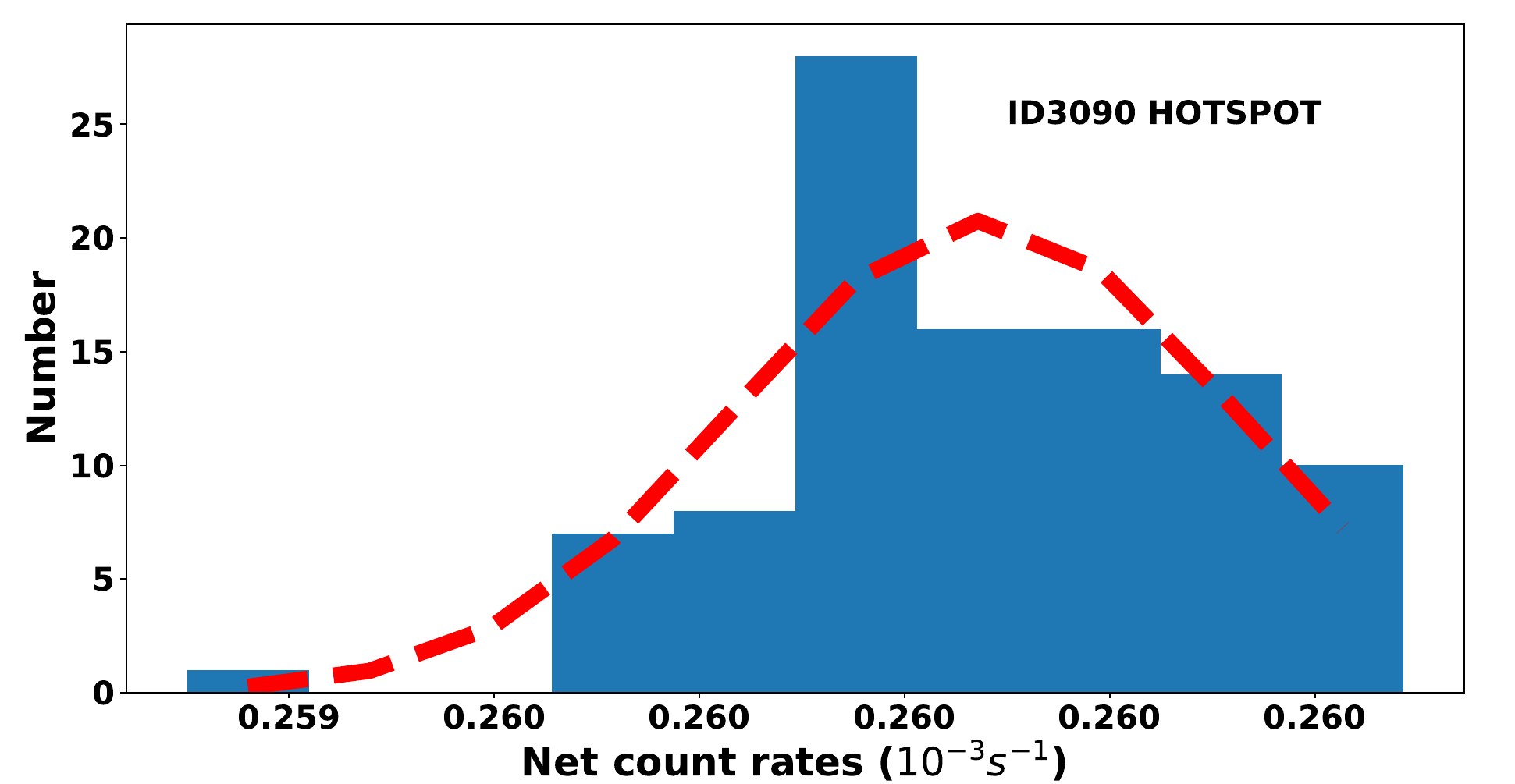}
\includegraphics[width=\columnwidth]{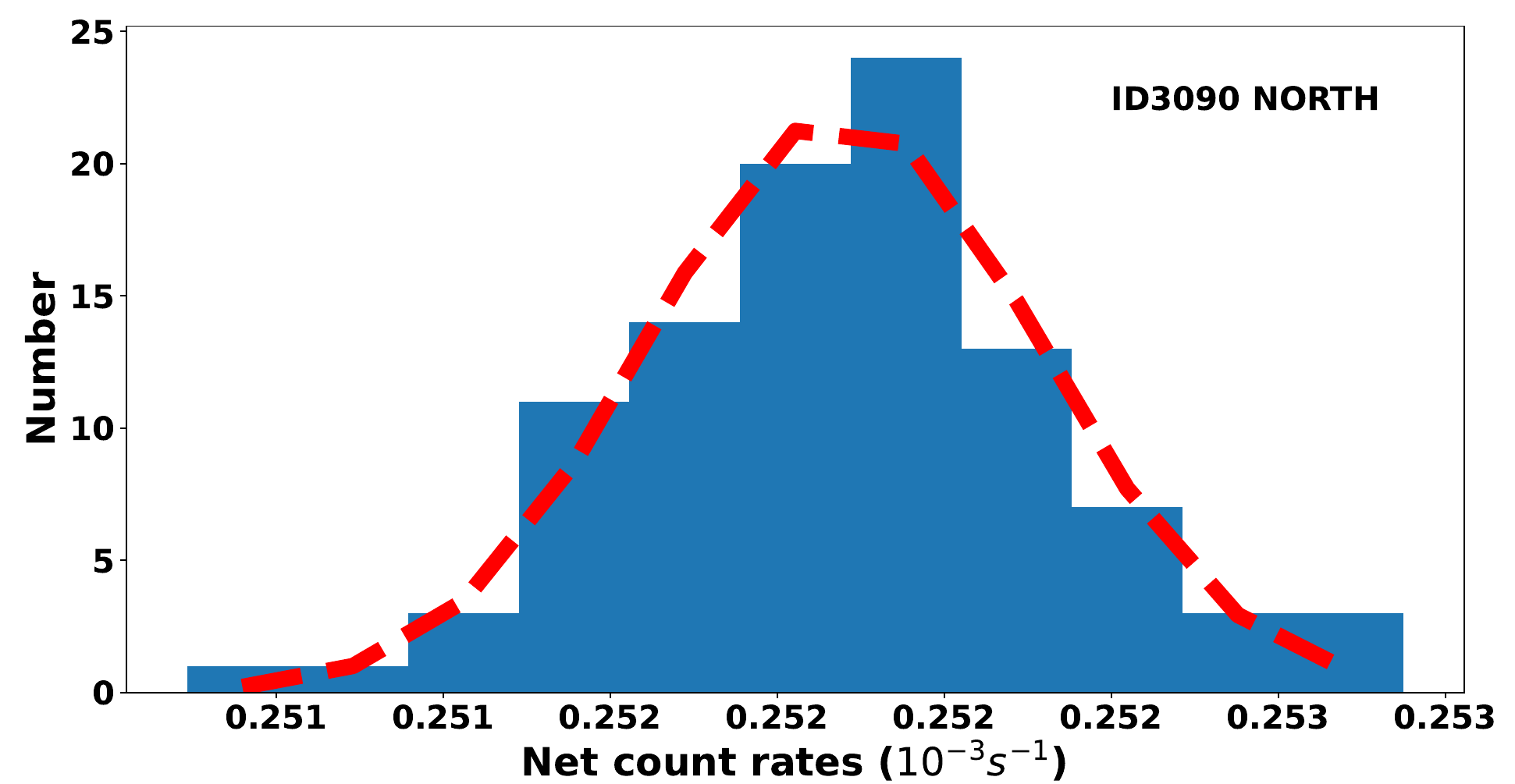}
\includegraphics[width=\columnwidth]{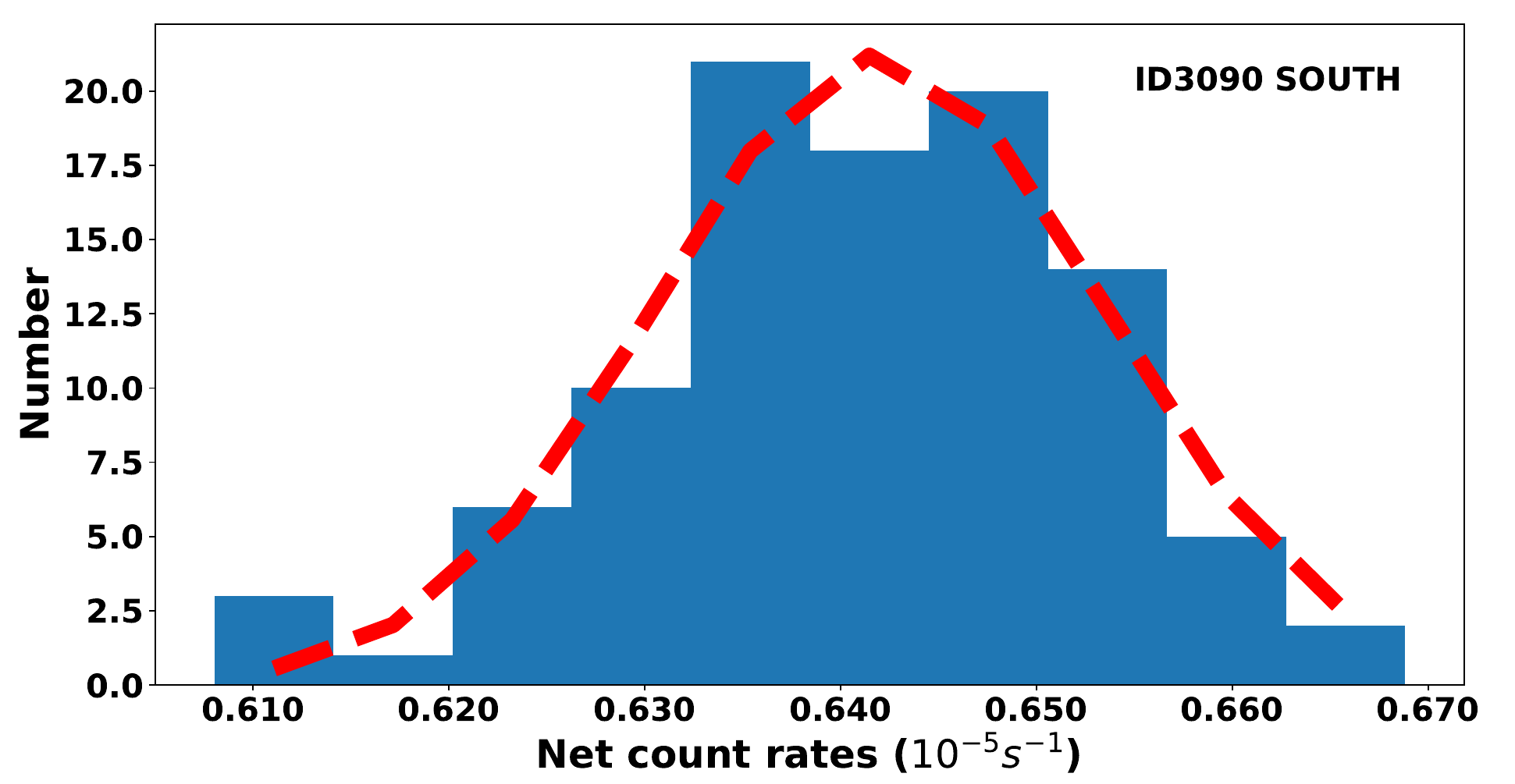}
\caption{Histograms of the net count rates calculated for the selected regions HS (top panel), N (middle panel), and S (lower panel) within the W hotspot in Pictor\,A, on the deconvolved exposure-corrected {\it Chandra} images for the ObsID 3090, for 100 random realizations of the PSF.}
\label{fig:histogram}
\end{figure}

\begin{figure}[!t]
\centering 
\includegraphics[width=\columnwidth]{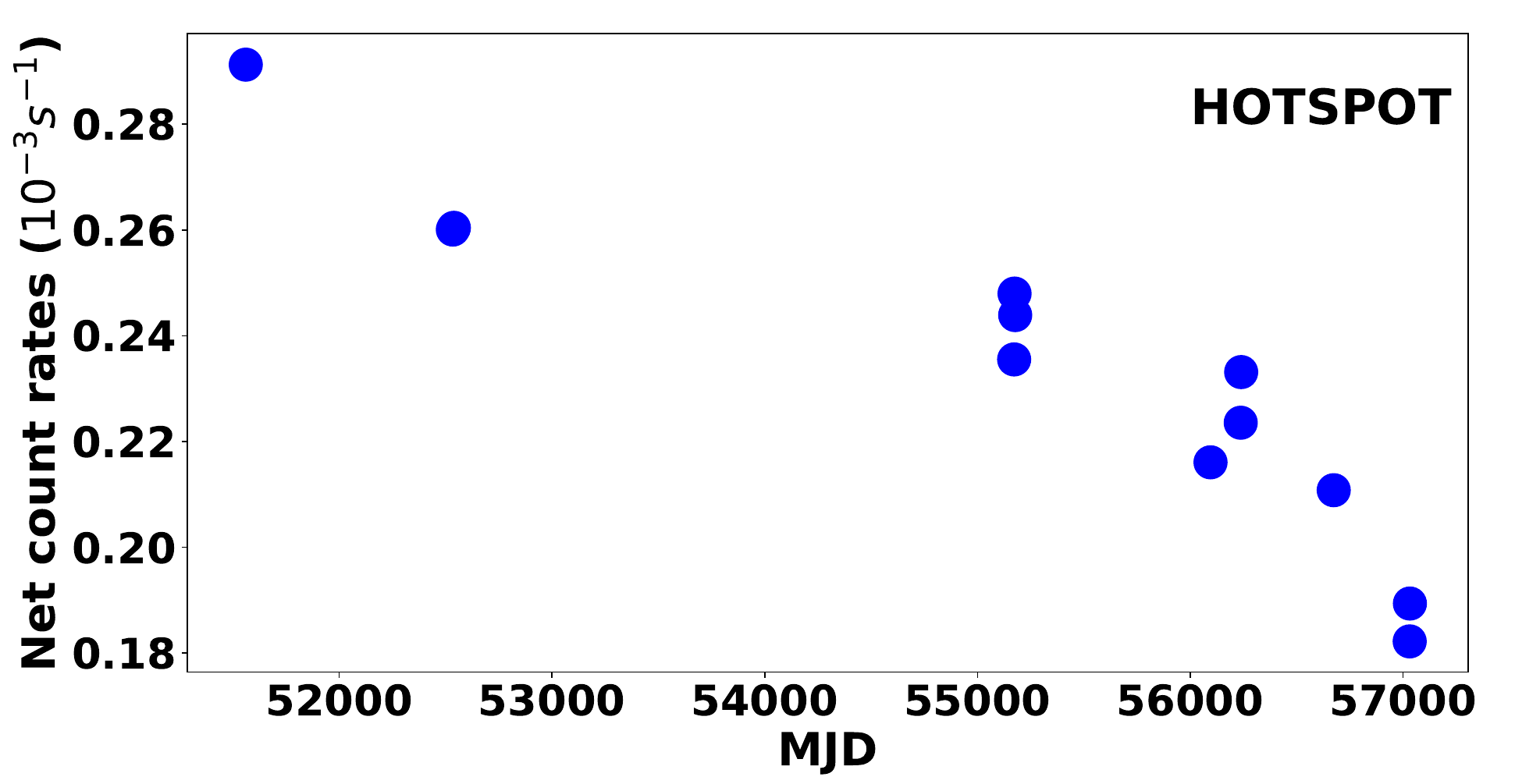}
\includegraphics[width=\columnwidth]{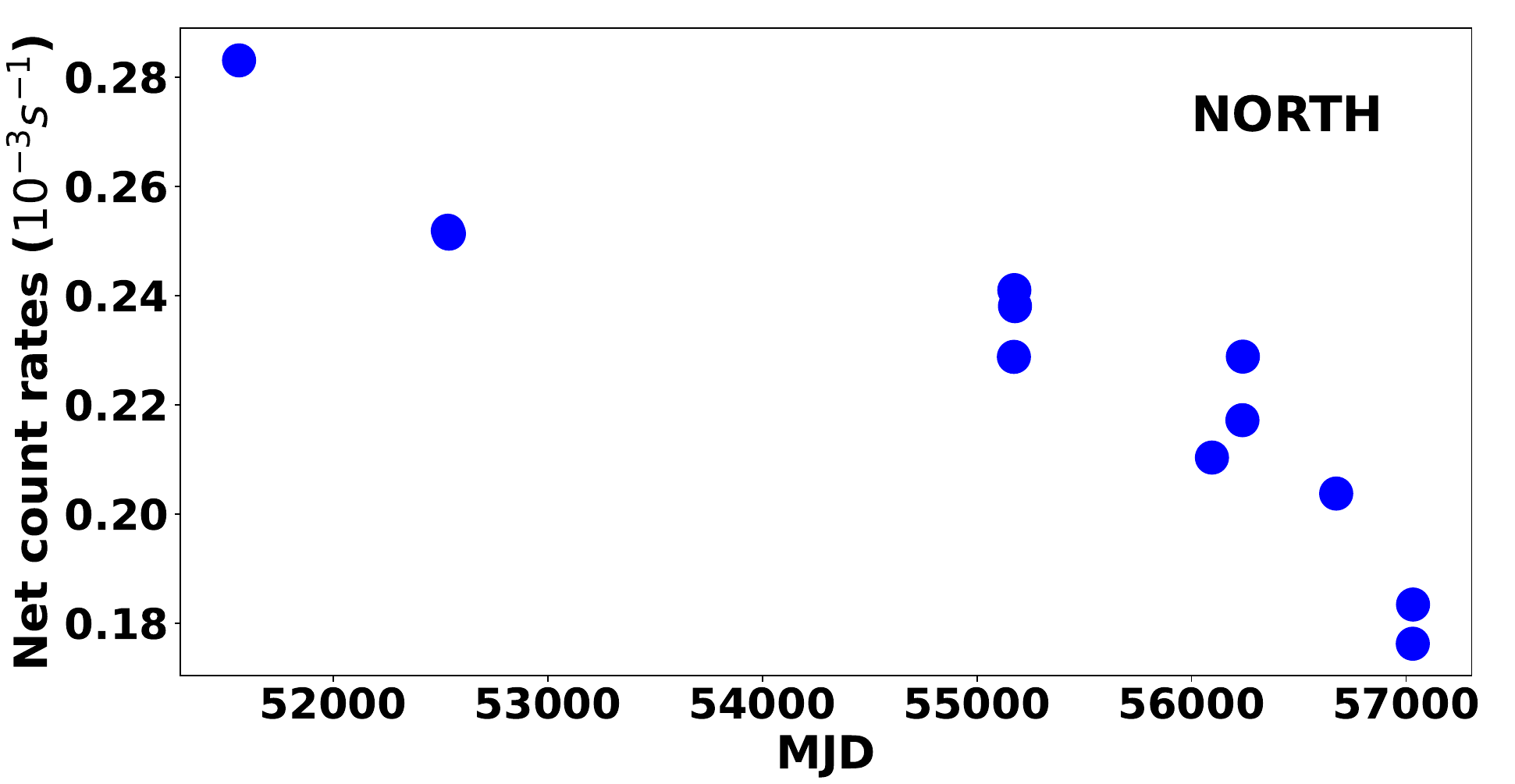}
\includegraphics[width=\columnwidth]{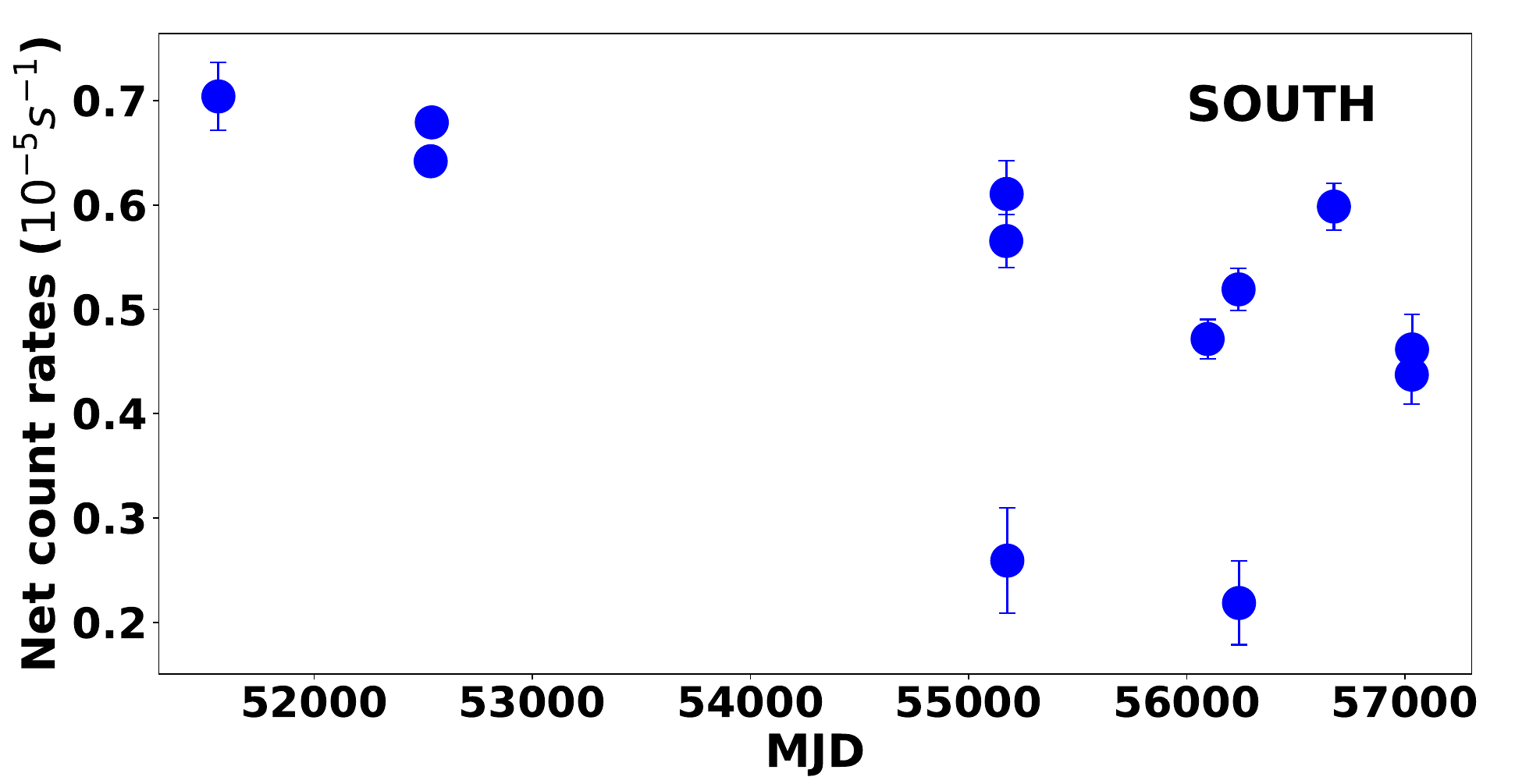}
\caption{The net count rate measured for the selected regions HS (top panel), N (middle panel), and S (lower panel) within the W hotspot in Pictor\,A on the deconvolved exposure-corrected {\it Chandra} images (see Table\,\ref{tab:rates}), as a function of the observing time.}
\label{fig:lightcurve}
\end{figure}

With such defined regions, we measure the net count rates (with the background chosen as adjacent to the hotspot but located outside of the radio lobe) individually on every image corresponding to particular random realizations of the PSF. As a result, for each given ObsID we obtained 100 independent net count rate measurements, for which the distribution is expected to be normal. Since the image deconvolution procedure does not provide any other way for propagating the errors, we consider the mean in those distributions $\mu$ as our final flux estimates, and the standard deviations, $\sigma$, as an estimate of the corresponding uncertainty in the performed flux measurements. Note in this context, that the PSF image is a simulated image based on a model, and there are uncertainties associated with the model adopted; by simulating 100 different realizations of the PSF we take into account the statistical uncertainties (so randomness of rays, detected rays, etc.), but not the systematic uncertainties in the model of the PSF simulator.

\begin{deluxetable}{ccccc}[!t]
\tabletypesize{\scriptsize}
\tablecaption{Net count rates of the W hotspot in Pictor\,A and its two sub-regions, from the deconvolved images.  \label{tab:rates}}
\tablehead{\colhead{ObsID} & \colhead{MJD} & \colhead{HS} & \colhead{N} & \colhead{S}\\
\colhead{} & \colhead{}  & \colhead{[$10^{-6}$\,cts\,s$^{-1}$]} & \colhead{[$10^{-6}$\,cts\,s$^{-1}$]} & \colhead{[$10^{-6}$\,cts\,s$^{-1}$]}}
\startdata
			346 & 51561 & 291.3$\pm$0.4 & 283.0$\pm$0.5 & 7.0$\pm$0.3\\					
			3090& 52534 & 260.1$\pm$0.2 & 251.8$\pm$0.3& 6.4$\pm$0.1\\
			4369&52539& 260.4$\pm$0.2& 251.3$\pm$0.4& 6.8$\pm$0.1\\
			12039&55172& 235.5$\pm$0.5& 228.7$\pm$0.7& 5.7$\pm$0.3\\
			12040&55174& 248.0$\pm$0.4& 241.0$\pm$0.6& 6.1$\pm$0.3\\
			11586&55177& 243.9$\pm$0.4& 238.0$\pm$1.0& 2.6$\pm$0.5\\
			14357&56095& 216.1$\pm$0.2& 210.3$\pm$0.4& 4.7$\pm$0.2\\
			14221&56237& 223.5$\pm$0.4& 217.1$\pm$0.5& 5.2$\pm$0.2\\
			15580&56239& 233.1$\pm$0.4& 228.8$\pm$0.7& 2.2$\pm$0.4\\
			14222&56674& 210.8$\pm$0.3& 203.7$\pm$0.4& 6.0$\pm$0.2\\
			16478&57031& 182.2$\pm$0.3& 176.2$\pm$0.5& 4.4$\pm$0.3\\
			17574&57032& 189.3$\pm$0.3& 183.4$\pm$0.4& 4.6$\pm$0.3
\enddata
\end{deluxetable}

Figure\,\ref{fig:histogram} presents the distributions of the net count rates for the ObsID 3090, along with the values of $\mu$ and $\sigma$ emerging from fitting Gaussians to the distributions. The flux estimates for all the analyzed ObsIDs, obtained in the analogous way by fitting Gaussians to the net count rate distributions, are summarized in Table\,\ref{tab:rates}; the corresponding histograms (HS regions only) are shown in the Appendix\,\ref{app:all}. Figure\,\ref{fig:lightcurve} presents the resulting lightcurves of the hotspot. The HS region is dominated by the N region, and in both, we observe a monotonic and statistically significant decrease of the net count rate by about $30\%$ between 2000 and 2015.

\begin{figure*}[!t]
\centering 
\includegraphics[width=0.48\textwidth]{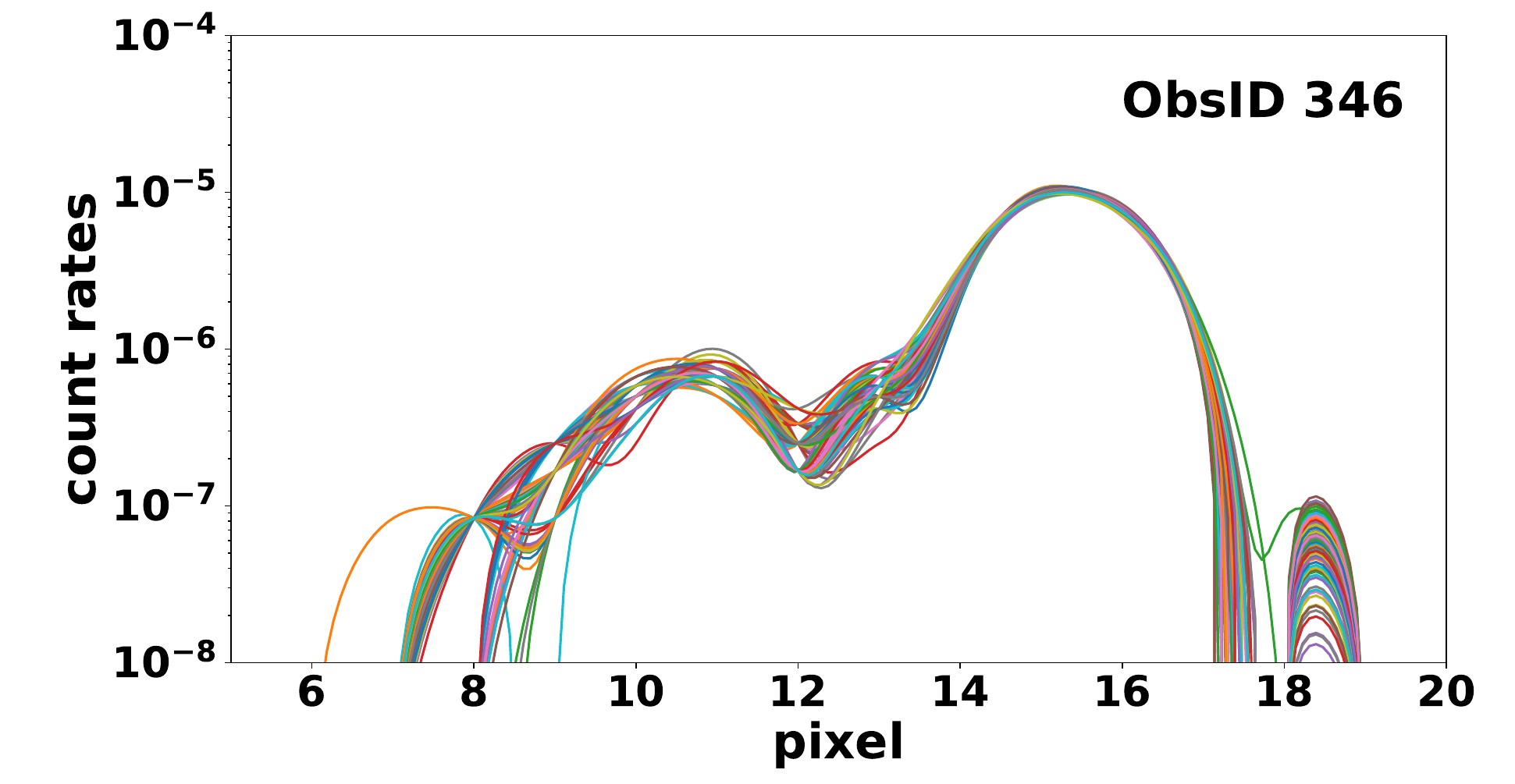}
\includegraphics[width=0.48\textwidth]{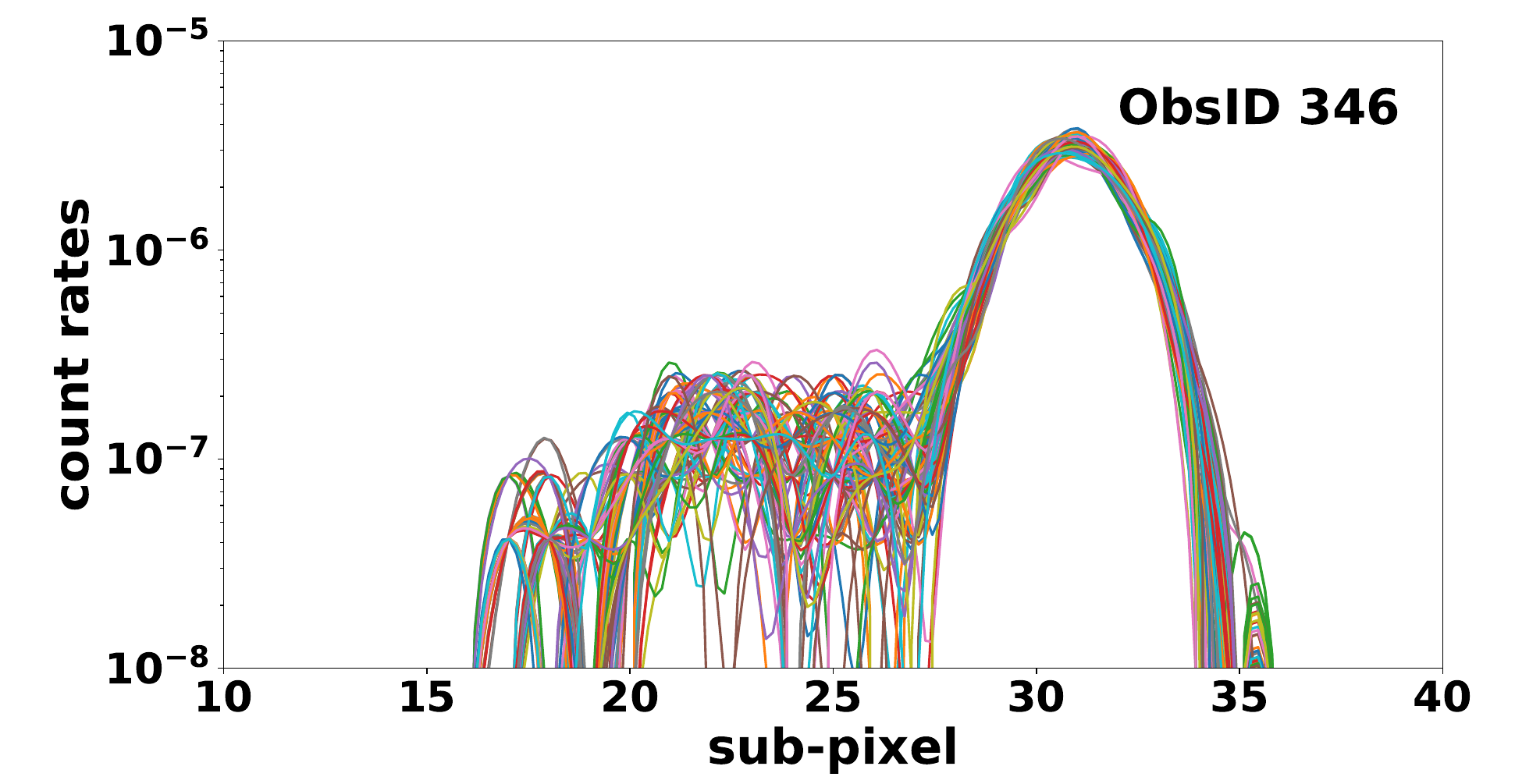}
\includegraphics[width=0.48\textwidth]{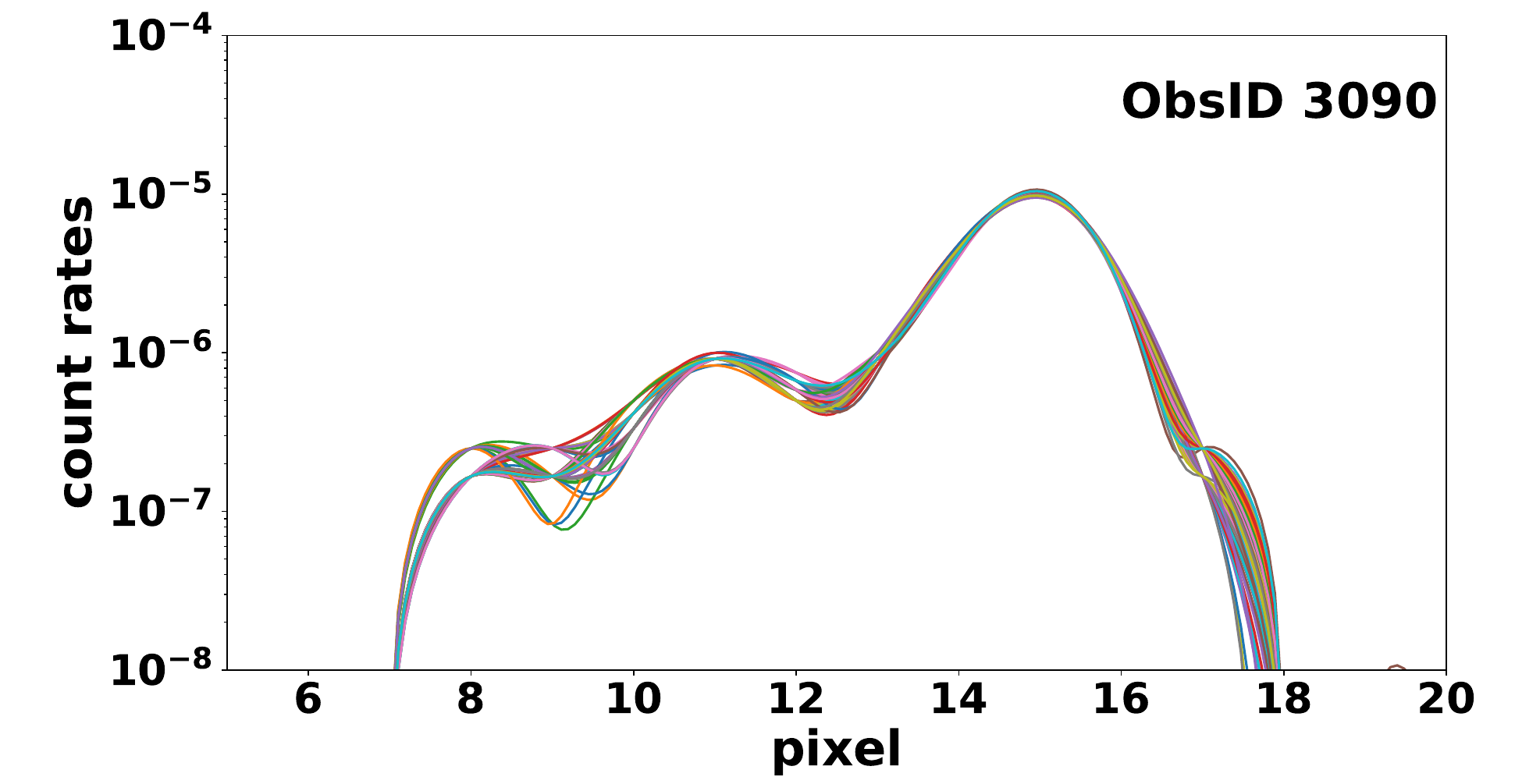}
\includegraphics[width=0.48\textwidth]{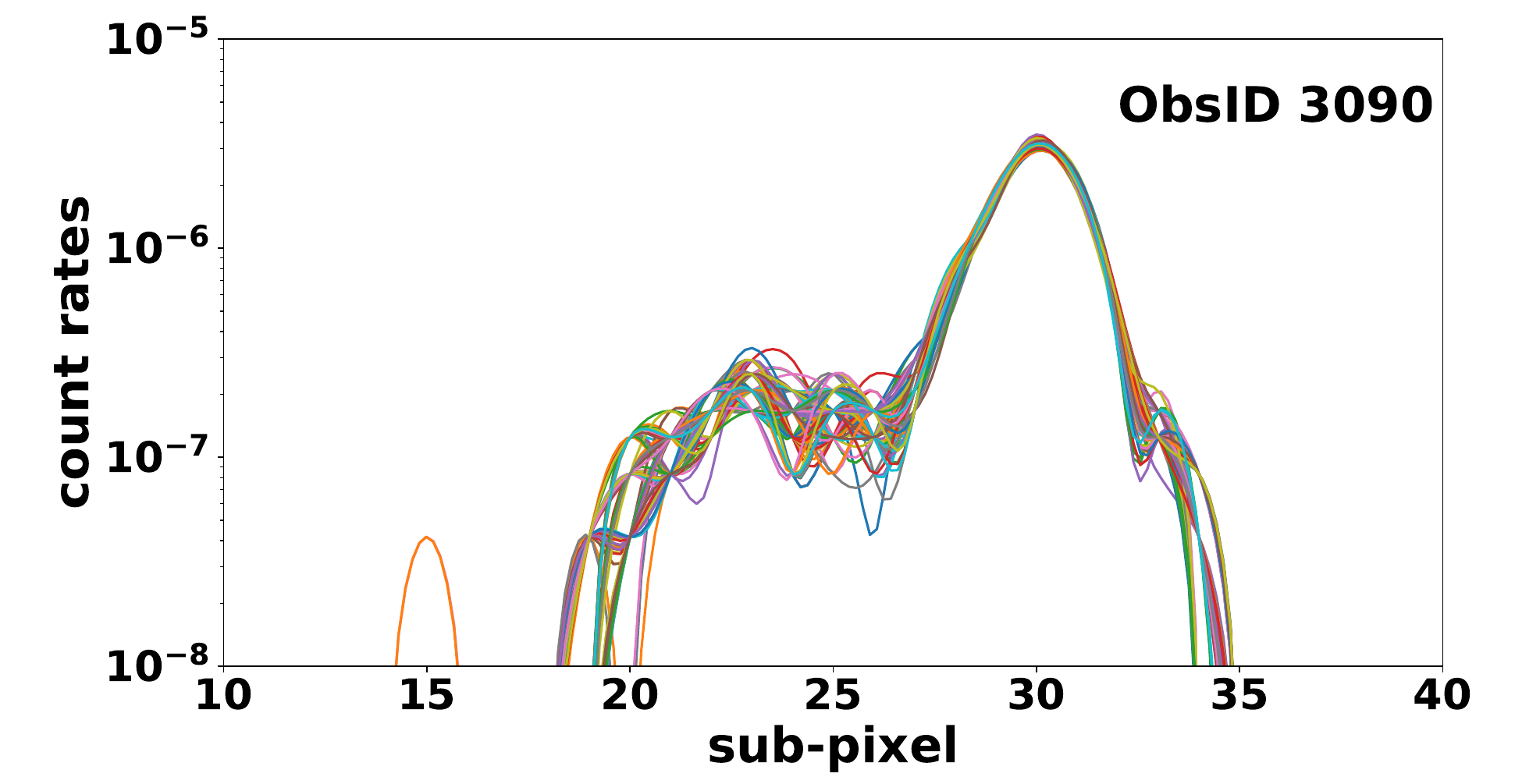}
\caption{Integrated intensity profiles along the major axis of the N region displayed on Figure\,\ref{fig:regions}, for the ObsID 346 and 3090 (upper and lower panels, respectively), at either 1\,px or 0.5\,px resolution (left and right panels, respectively). Each curve on the panels corresponds to a single realization of the PSF for a given ObsID.}
\label{fig:profile}
\end{figure*}

As the deconvolution procedure was performed on the exposure-corrected images, we do not believe that this decrease could be due to the known degradation of the ACIS CCDs \citep[see][]{Plucinsky18}, at least not entirely --- some of the flux changes have to be intrinsic to the hotspot, especially after 2010 when the observed net count rate decrease seemingly sped up. Interestingly, in the S region we also see the overall net count rate decrease between 2000 and 2015, and this could be considered as evidence for the observed flux changes being rather due to the CCDs' degradation. On the other hand, for the S region the lightcurve seems much more erratic, and the overall net count rate is at the level of a few percent of the net count rate for the N region, and so it is possible that what we observe here is rather a photon leakage from the N region to the adjacent S region.

If the observed X-ray flux changes, at the level of $30\%$, are indeed intrinsic to the hotspot, they have to originate within the brightest part of the source, i.e. within the Mach disk/conical shock. This feature is clearly extended for about $4^{\prime\prime}$ in the longitudinal direction (i.e., perpendicularly to the jet axis), meaning a physical scale of about 3\,kpc; the question is if it can also be resolved transversely. In order to investigate this issue, we use the {\fontfamily{qcr}\selectfont dmregrid2} tool implemented in the {\fontfamily{qcr}\selectfont CIAO} to rotate the deconvolved {\it Chandra} images of the hotspot, and we extract the integrated counts along the major axis of the N region displayed on Figure\,\ref{fig:regions}. We perform this exercise separately for each realization of the PSF for a given ObsID. As a result, we obtain 100 surface brightness profiles for each ObsID. These are shown for comparison for the ObsIDs 346 and 3090 on Figure\,\ref{fig:profile} (left and right panels, respectively), at either 1\,px or 0.5\,px resolution (upper and lower panels, respectively). The profiles for all the analyzed ObsIDs (only sub-px resolution) are displayed in the Appendix\,\ref{app:all} below. As follows, on the 0.5\,px-resolution de-convolved images the brightest segment of the hotspot appears narrower than on the 1\,px-resolution de-convolved images. The intensity profiles of this segment are in all the cases (except of ObsID 14357 commented above in \S\,\ref{sec:deconvolve}) symmetric and centrally-peaked; this indicates that the shock structure is transversely unresolved even at the sub-px {\it Chandra} resolution, with the corresponding scale upper limit of $\sim 0.^{\prime\prime}25 < 200$\,pc.

\section{Conclusions} 
\label{sec:conclusions}

The Western hotspot in the radio galaxy Pictor\,A is known for its complex multi-wavelength morphology, and also for its broad-band emission spectrum defying simple modeling. In the context of the former, one may reiterate on the presence of the extended radio/optical filament perpendicular to the jet axis and located several kpc upstream of the termination shock \citep{Perley97,Thomson95,Saxton02}, as well as of the high-brightness temperature compact (pc-scale) radio knots within/around the termination shock \citep{Tingay08}. Regarding the latter, one may note the mid-infrared excess over the integrated radio-to-optical continuum of the entire structure \citep{Isobe17}, as well as the intense power-law X-ray emission, for which both the observed flux and the best-fit slope challenge all simple ``one-zone'' emission scenarios, ascribing the production of the observed keV-energy photons to either synchrotron or inverse-Compton processes \citep[e.g.,][]{Wilson01}. 

Multiple {\it Chandra} observations of the source, summarized and re-analyzed in \citet{H16}, enriched the overall picture by a tentative detection of the X-ray flux variability within the hotspot, on the relatively short timescale of years.

The novelty of the analysis of the {\it Chandra} data presented here, is that by means of detailed PSF simulations and image deconvolution, we were able to resolve the X-ray structure of the hotspot into (i) the jet-like feature located in between the radio/optical filament and the termination shock, and (ii) the disk-like or conical feature perpendicular to the jet axis, and located $\sim 1.^{\prime\prime}5 \simeq 1$\,kpc upstream the intensity peak of the radio hotspot. We believe that this later feature --- resolved in its longitudinal direction to be $\sim 4^{\prime\prime} \sim 3$\,kpc long, while remaining basically unresolved in its transverse direction, with the corresponding scale upper limit of $\sim 0.^{\prime\prime}25 < 200$\,pc --- marks the position of the reverse shock front in the system, where efficient particle acceleration takes place. Note in this context, that in the case of the reverse shock, one is dealing with mildly-relativistic plasma bulk velocities \citep{Meisenheimer89,Kino04}, and a quasi-perpendicular magnetic configuration as evidenced by the radio polarimetry of the hotspot \citep{Perley97}.

We also noted the monotonically decreasing count rate on the deconvoled {\it Chandra} images, amounting to about $\sim 30\%$ drop over the 15 years covered by the {\it Chandra} monitoring (January 2000 -- January 2015). We believe that this decrease cannot be explained as (solely) due to the degradation of the ACIS CCDs.

The finding that the transverse size of the shock front at X-ray frequencies turns out to be less than 200\,pc, should not be surprising, however, once the X-ray production mechanism is identified as a synchrotron process. Indeed, the observed propagation length of ultra-relativistic electrons emitting synchrotron photons at a given (observed) frequency $\nu_{\rm syn}$, is $\ell_{\rm rad} \simeq \beta c \Gamma \times \tau'_{\rm rad}\!(\nu_{\rm syn})$, where $\beta$, $\Gamma$, and $\delta$ are the bulk velocity, the bulk Lorentz factor, and the bulk Doppler of the emitting plasma, respectively, while $\tau'_{\rm rad}\!(\nu_{\rm syn})$ is the radiative cooling timescale, as measured in the plasma rest frame for the electrons with Lorentz factors corresponding to the frequency of the synchrotron photons $\gamma \propto \sqrt{\nu_{\rm syn} / \delta B}$; in particular
\begin{equation}
\left(\frac{\ell_{\rm rad}}{\rm pc} \right) \sim \frac{10 \, \beta \Gamma \delta^{1/2} B_{\rm 100\,\mu G}^{-3/2}}{1+10^{-3} \Gamma^2 B_{\rm 100\,\mu G}^{-2} } \times \left(\frac{\nu_{\rm syn}}{10^{18}\,{\rm Hz}}\right)^{-1/2}
\label{eq:rad}
\end{equation}
where $B_{\rm 100\,\mu G}$ is the magnetic field intensity within the emission region in the units of $100$\,$\mu$G, and we also included the inevitable radiative losses related to the inverse-Compton up-scattering of the Cosmic Microwave Background photons \citep[at the source redshift of $z \simeq 0.035$; see][equation 7 therein]{Stawarz04}. 

The above equation, with the expected $\beta \sim 0.3$ characterizing the downstream of the reverse shock \citep[e.g.,][]{Meisenheimer89,Kino04}, and so $\Gamma \sim \delta \sim 1$, as well as $B_{\rm 100\,\mu G} \gtrsim 1$ \citep[see][]{Meisenheimer89,Meisenheimer97,Isobe17}, gives $\ell_{\rm rad} \lesssim 3$\,pc for keV energies of the observed synchrotron photons. Interestingly, such a small spatial scale would be in rough agreement with the observed variability timescale as well, since the corresponding light-crossing timescale $\ell_{\rm rad}/c \lesssim 10$\,yr.

The simple estimates presented above seem, therefore, to suggest that an efficient acceleration of ultra-relativistic electrons, up to energies enabling the production of synchrotron X-ray photons ($\gamma \sim 10^7$), takes place exclusively within a rather thin layer of the very front of the termination shock in the radio galaxy Pictor\,A. In this framework, the fact that the radio intensity peak position is located further downstream, could be explained as solely due to the increasing volume of the emitting plasma around the position of the contact discontinuity, where the flow (i.e., shocked jet material) diverges. Interestingly, for the radio-emitting electrons (say, $\nu_{\rm syn} \simeq 10$\,GHz), the above equation \ref{eq:rad} returns $\ell_{\rm rad} \sim 30 B_{\rm 100\,\mu G}^{-3/2}$\,kpc, which would be consistent with the observed X-ray/radio offset of the order of $\sim 1$\,kpc, if only the magnetic field intensity within the hotspot (downstream of the reverse shock) is $B \sim 1$\,mG, i.e. a factor of a few above the equipartition value. Moreover, for electrons emitting optical synchrotron photons in $\sim 1$\,mG magnetic field, the observed propagation length reads as $\ell_{\rm rad} \sim 10$\,pc, in agreement with no pronounced offset observed between the positions of the X-ray and optical intensity peaks. Still, in such a case, the presence of the upstream X-ray enhancement (i.e., the jet-like feature located in between the radio/optical filament and the termination shock), would remain unexplained. Hydrodynamical simulations of light, supersonic jets propagating within hot gaseous atmospheres, such as those presented by, e.g., \citet{Saxton02} or \citet{Mizuta04}, hint however a rather complex morphology around the jet termination regions, consisting of a network of various shocks, backflows, and vortex structures, and so any robust identification of the features revealed by our image deconvolution awaits a more careful comparison with the numerical data.

\acknowledgments

This research has made use of data obtained from the Chandra Data Archive. This work was supported by the Polish NSC grant 2016/22/E/ST9/00061 (R.T., \L .S., V.M., K.B.). Work at the Naval Research Laboratory is supported by the Chief of Naval Research. A.S. was supported by NASA contract NAS8-03060 (Chandra X-ray Center).
\vspace{5mm}
\facilities{Chandra (ACIS)}
\software{{\fontfamily{qcr} \selectfont CIAO} \citep{Fruscione06}, {\fontfamily{qcr}\selectfont Sherpa} \citep{Freeman01}, {\fontfamily{qcr} \selectfont ChaRT} \citep{Carter03}, and {\fontfamily{qcr} \selectfont MARX} \citep{Davis12} } 
   

\clearpage

\appendix
\section{Net count rates and integrated intensity profiles for all the Chandra pointings}
\label{app:all}

\begin{figure*}[h]
\centering 
\includegraphics[width=0.24\textwidth]{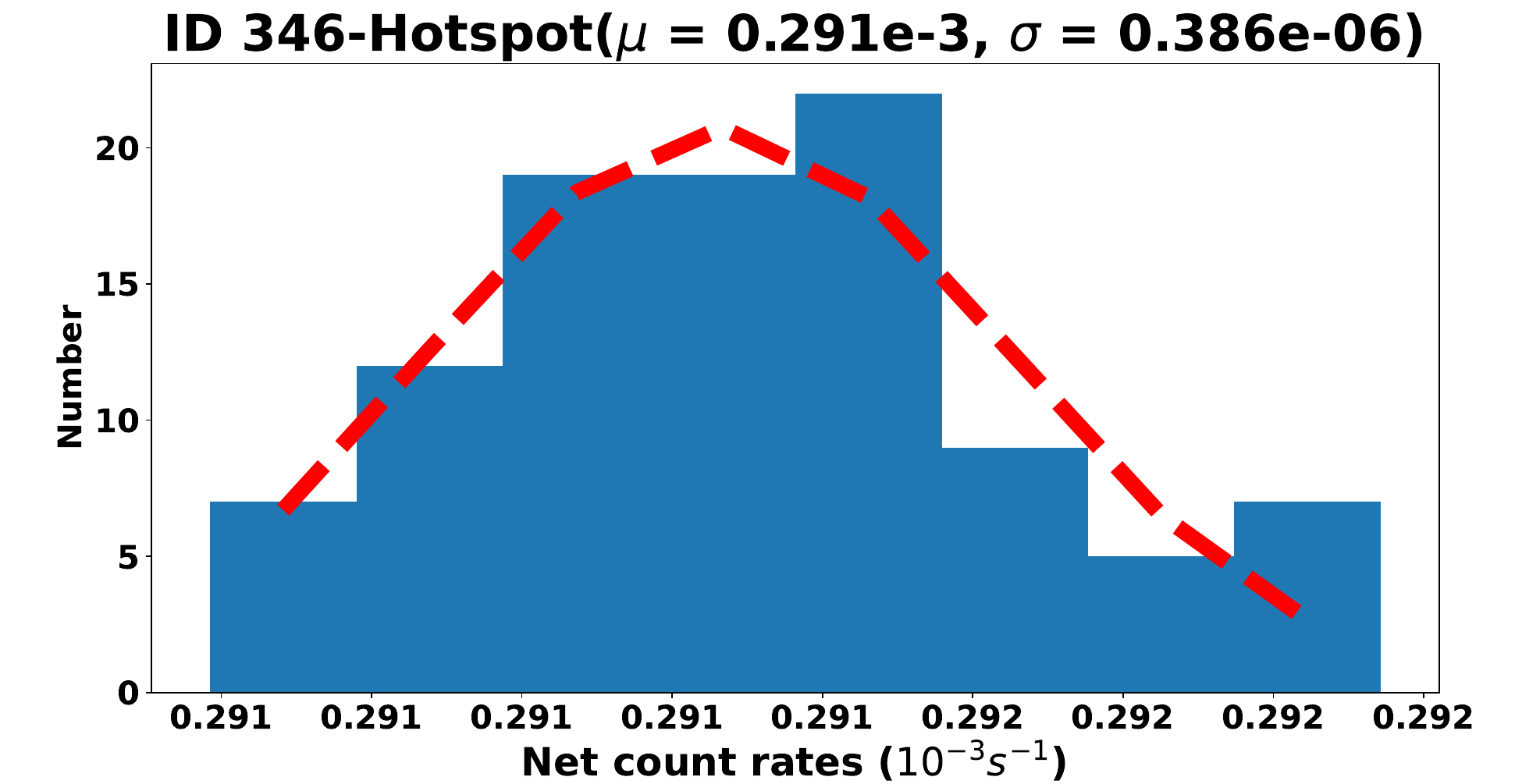}
\includegraphics[width=0.24\textwidth]{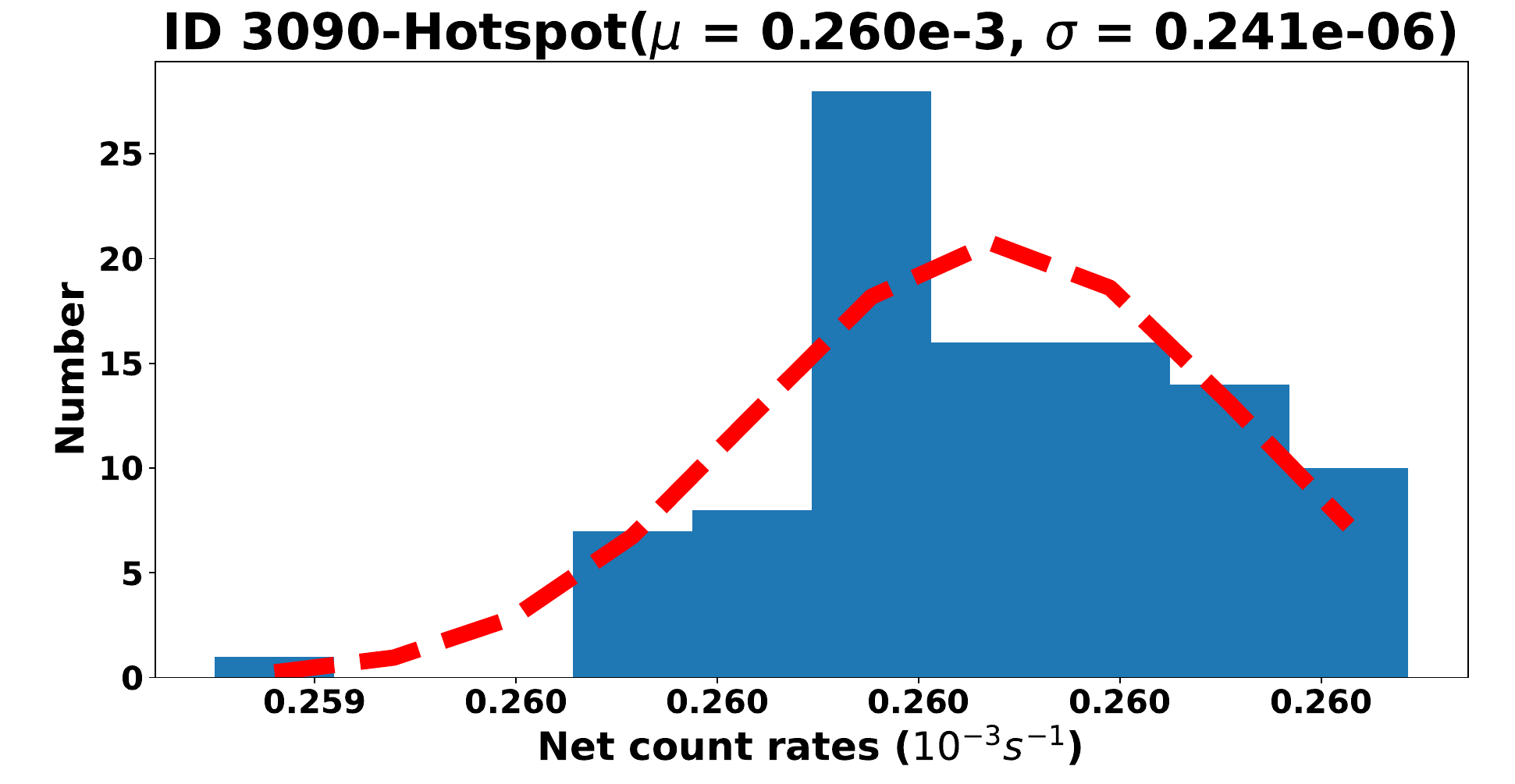}
\includegraphics[width=0.24\textwidth]{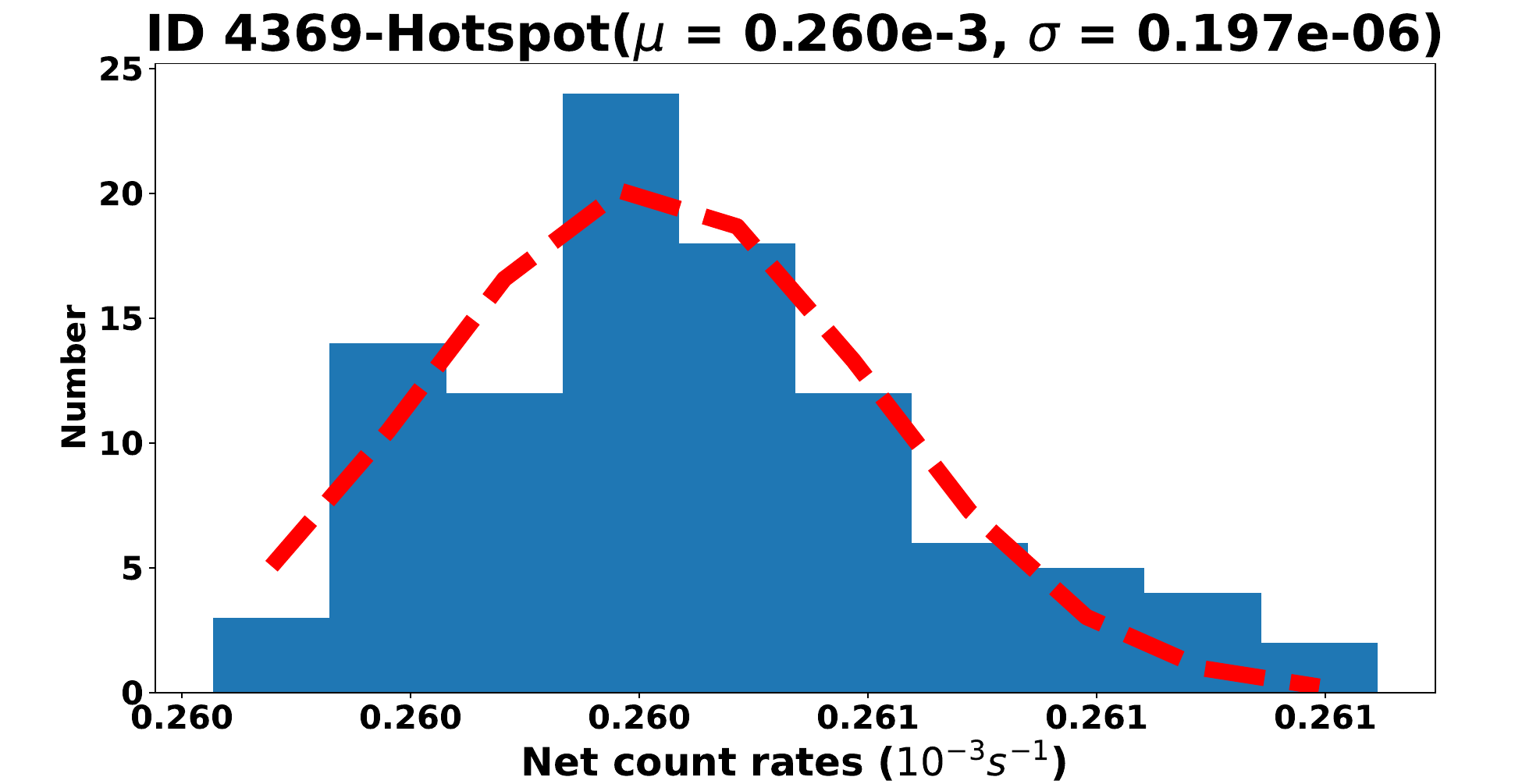}
\includegraphics[width=0.24\textwidth]{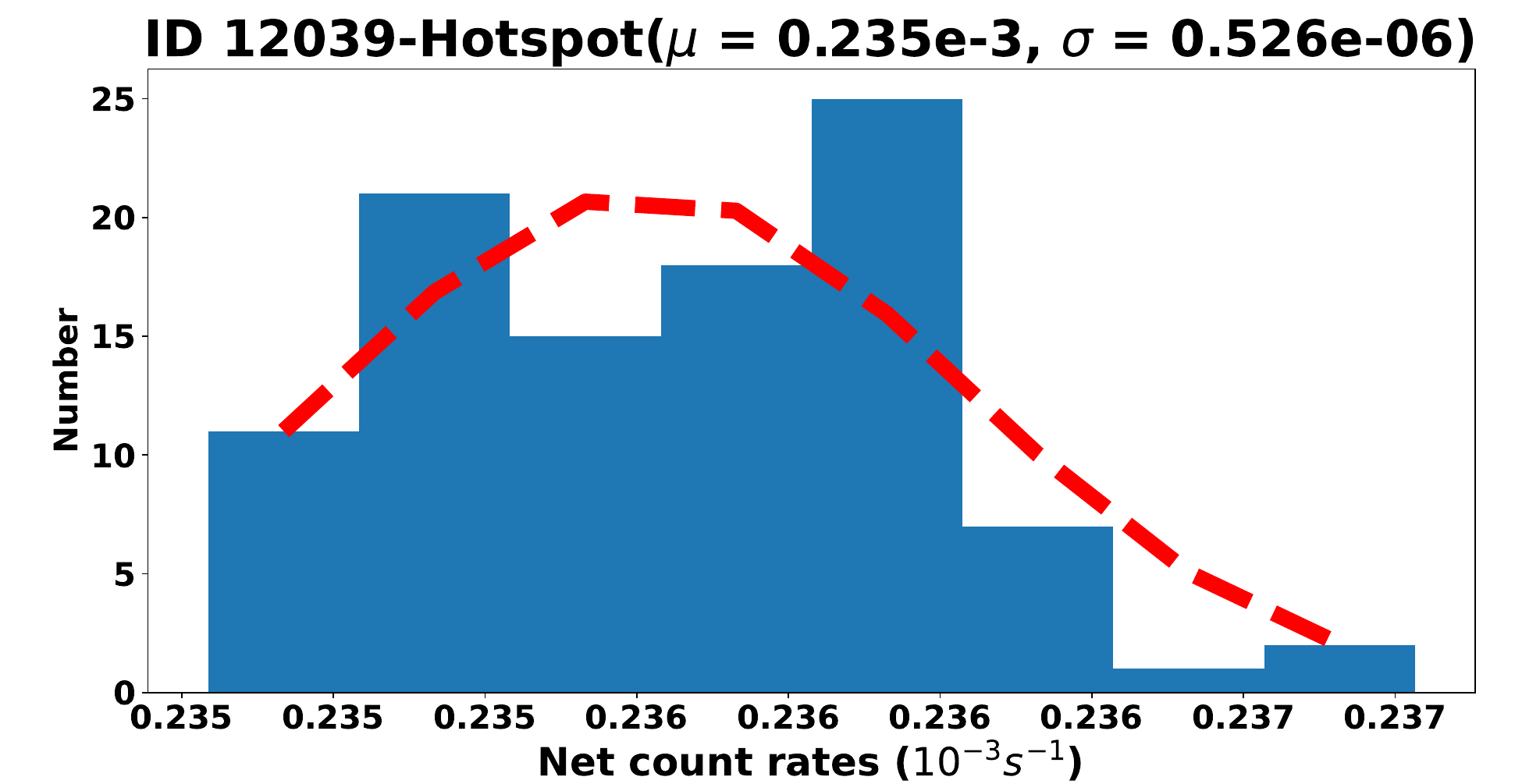}
\includegraphics[width=0.24\textwidth]{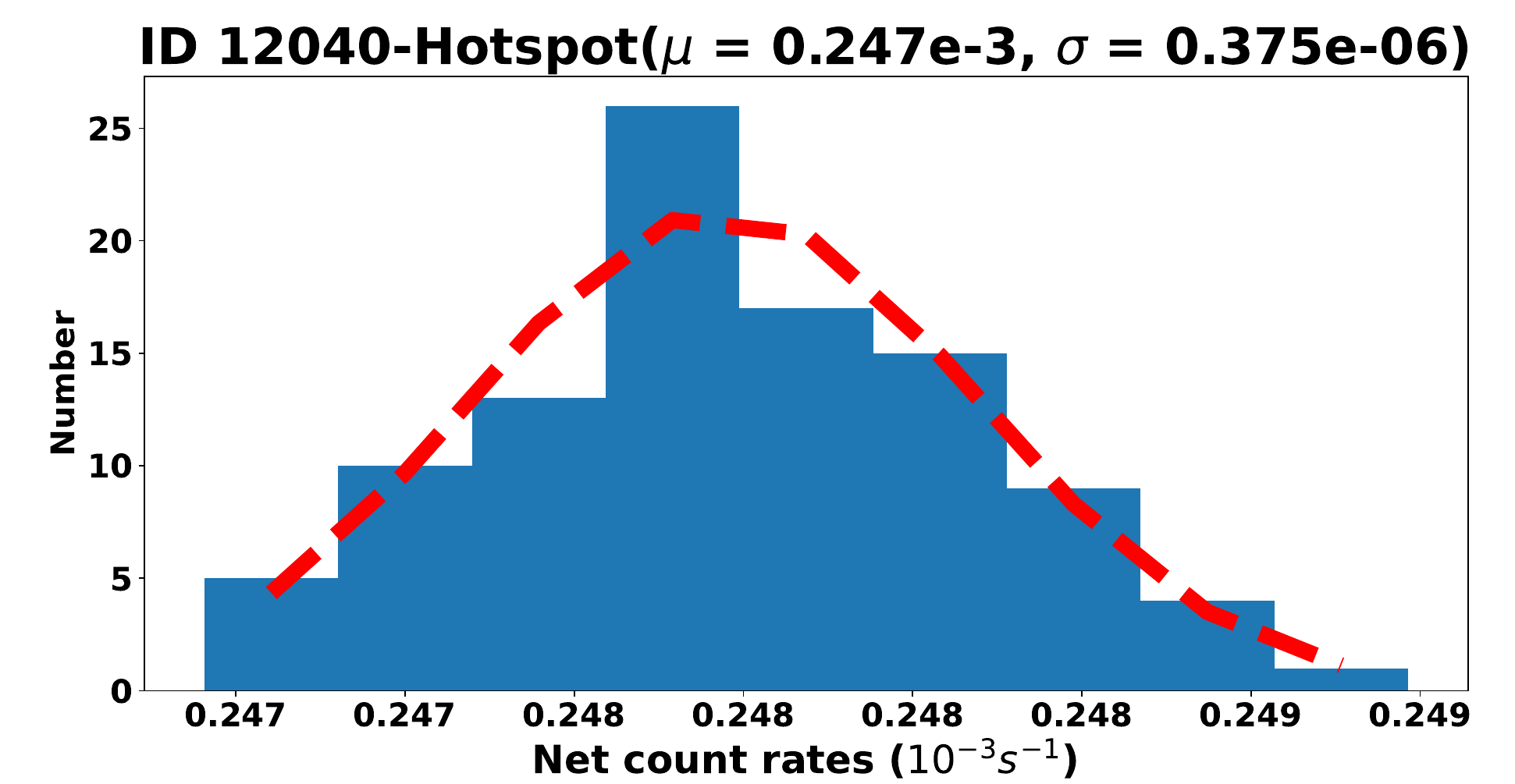}
\includegraphics[width=0.24\textwidth]{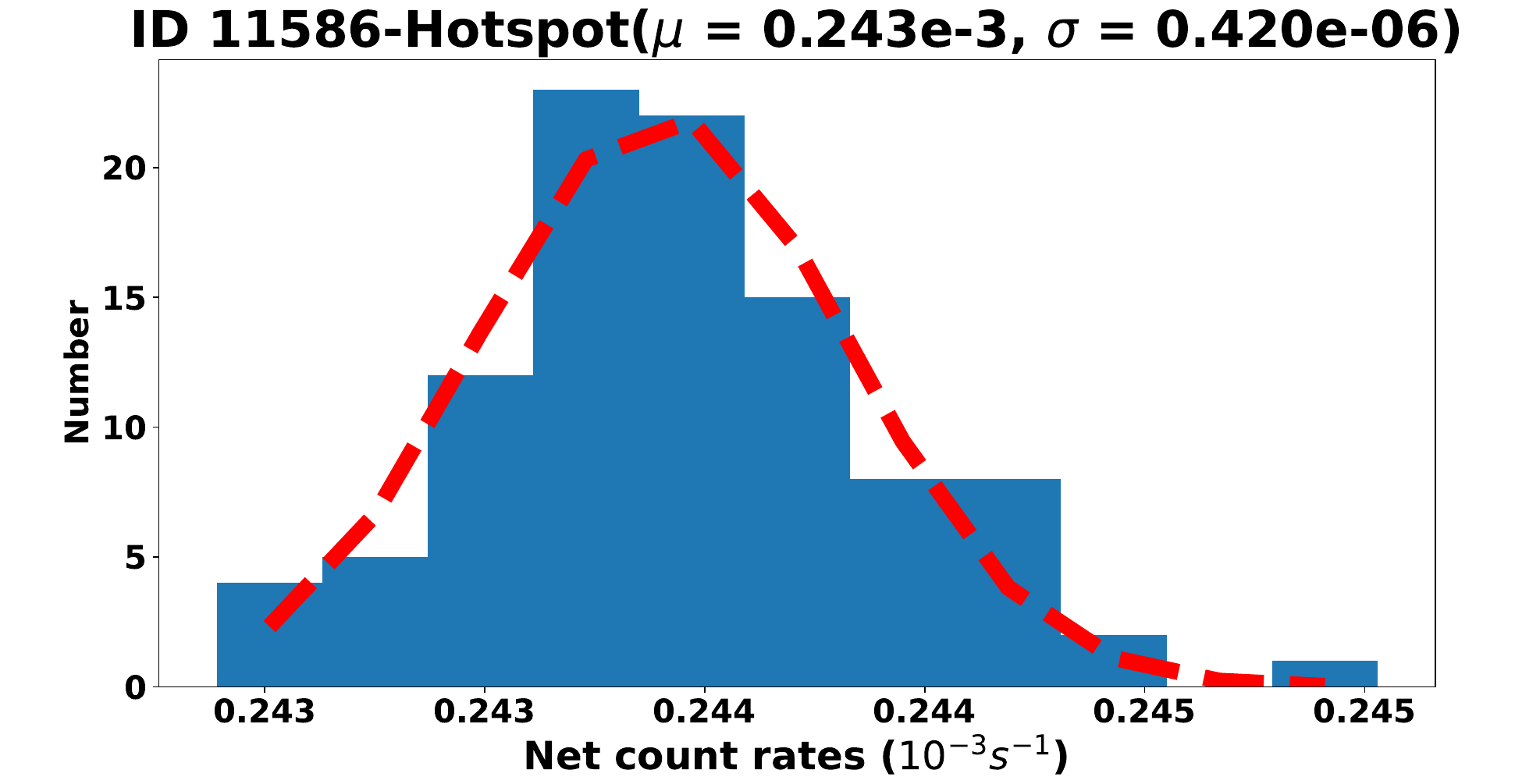}
\includegraphics[width=0.24\textwidth]{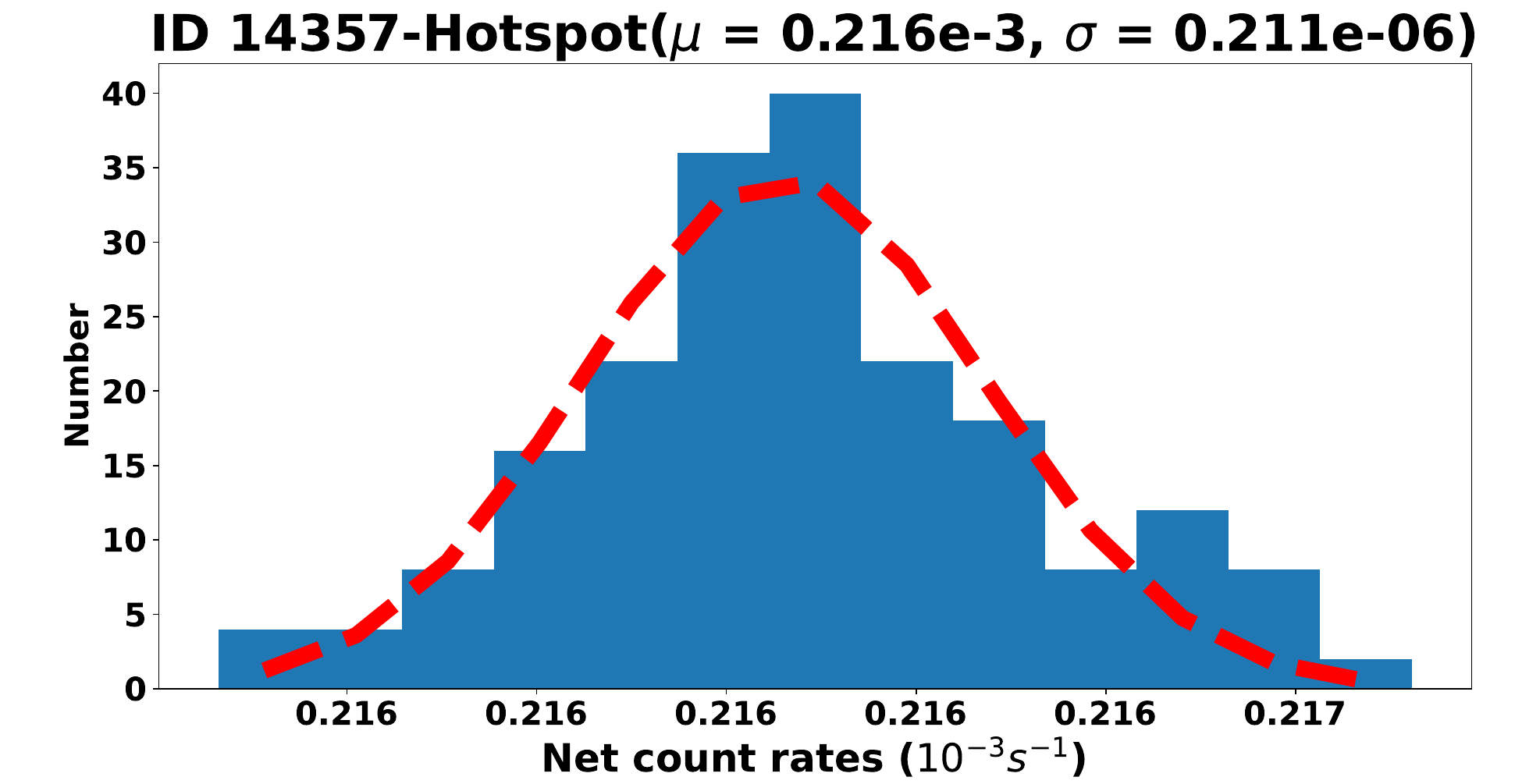}
\includegraphics[width=0.24\textwidth]{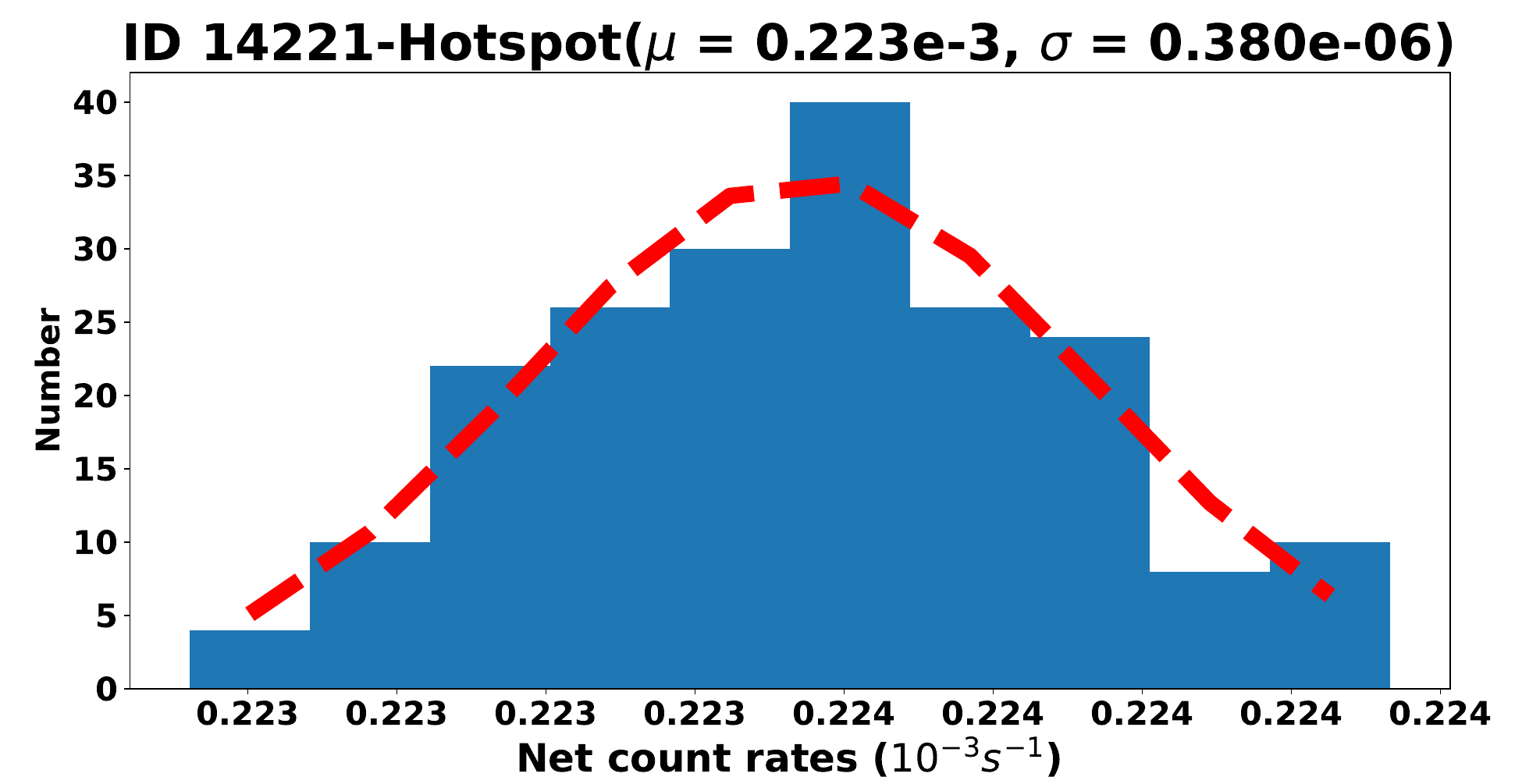}
\includegraphics[width=0.24\textwidth]{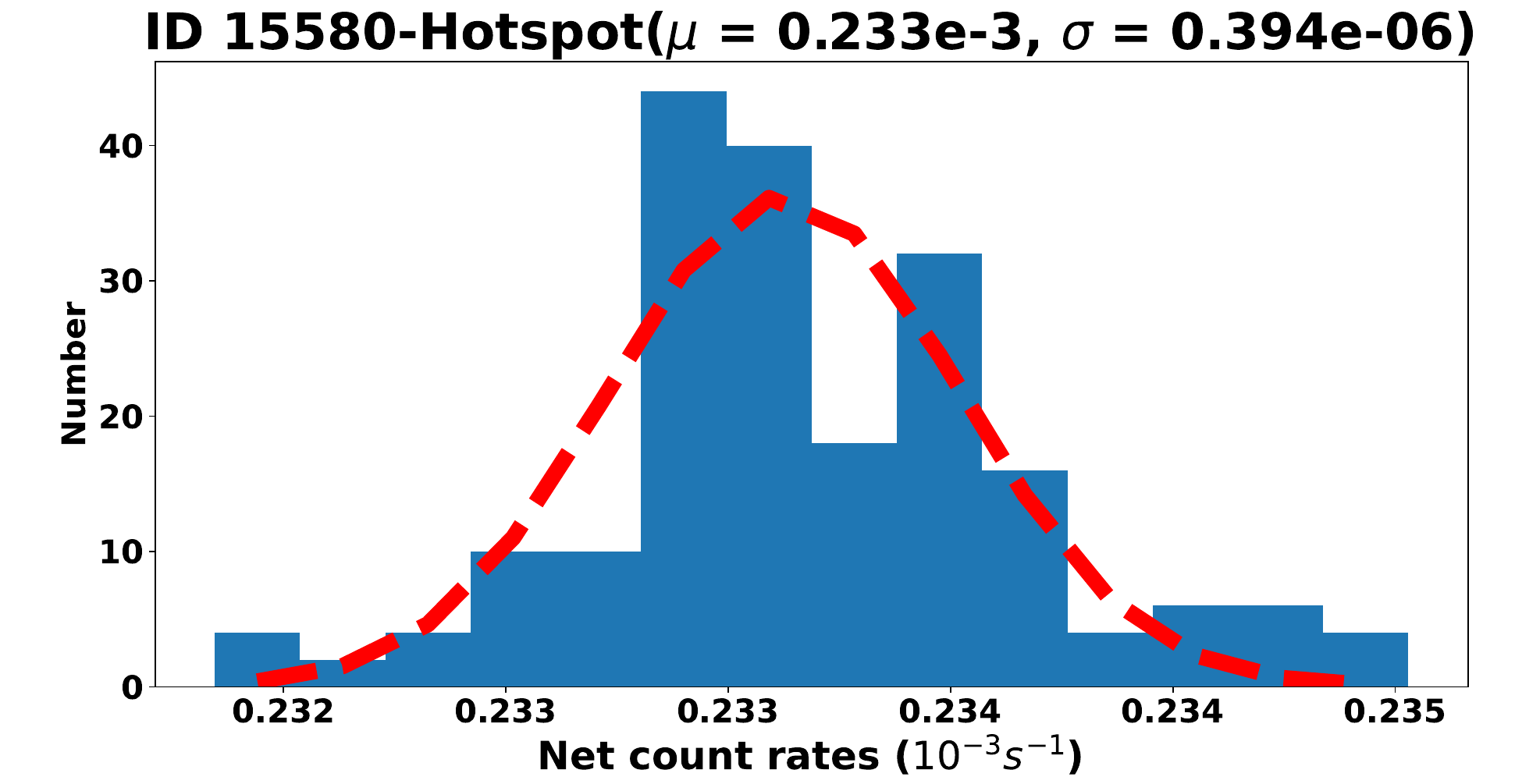}
\includegraphics[width=0.24\textwidth]{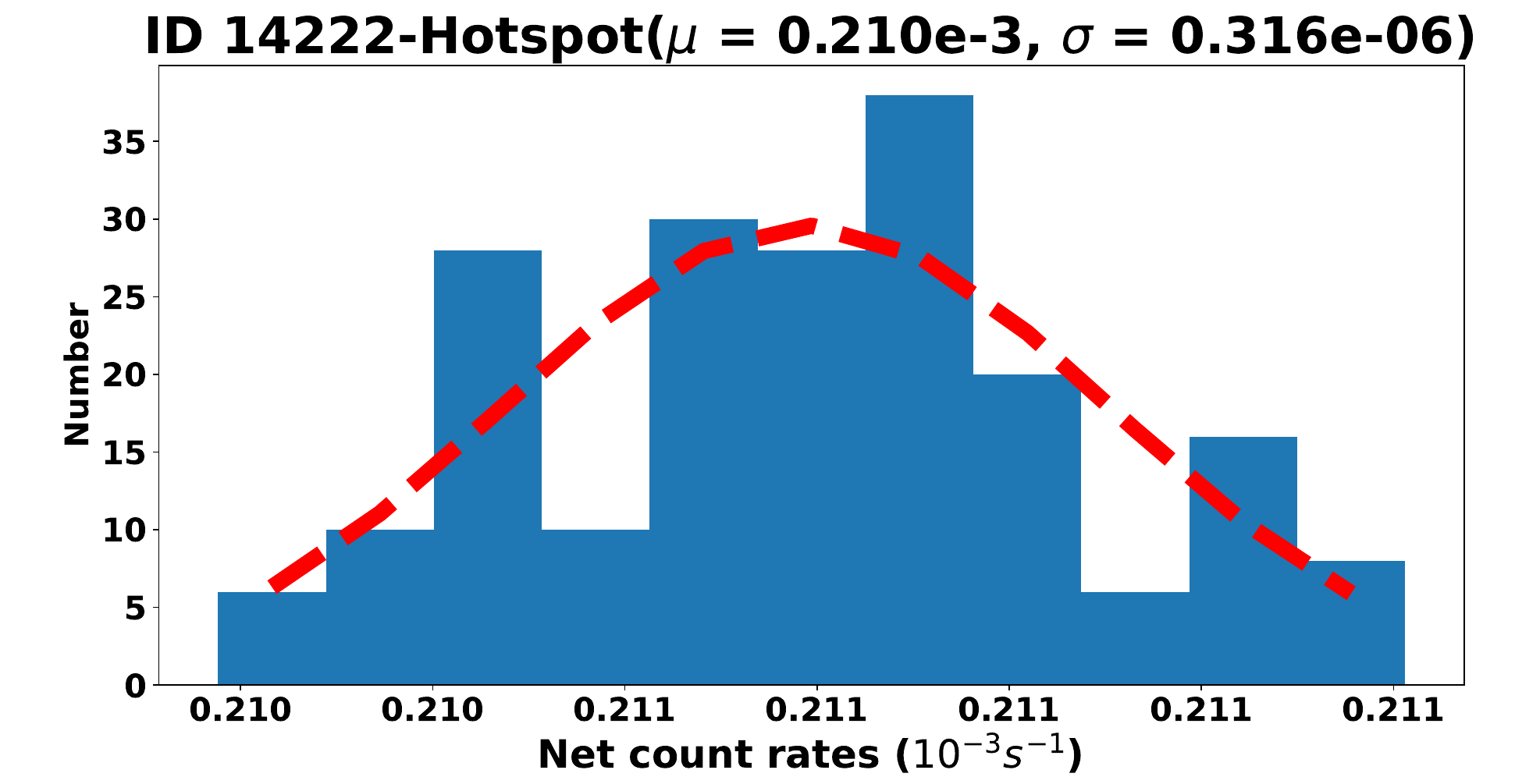}
\includegraphics[width=0.24\textwidth]{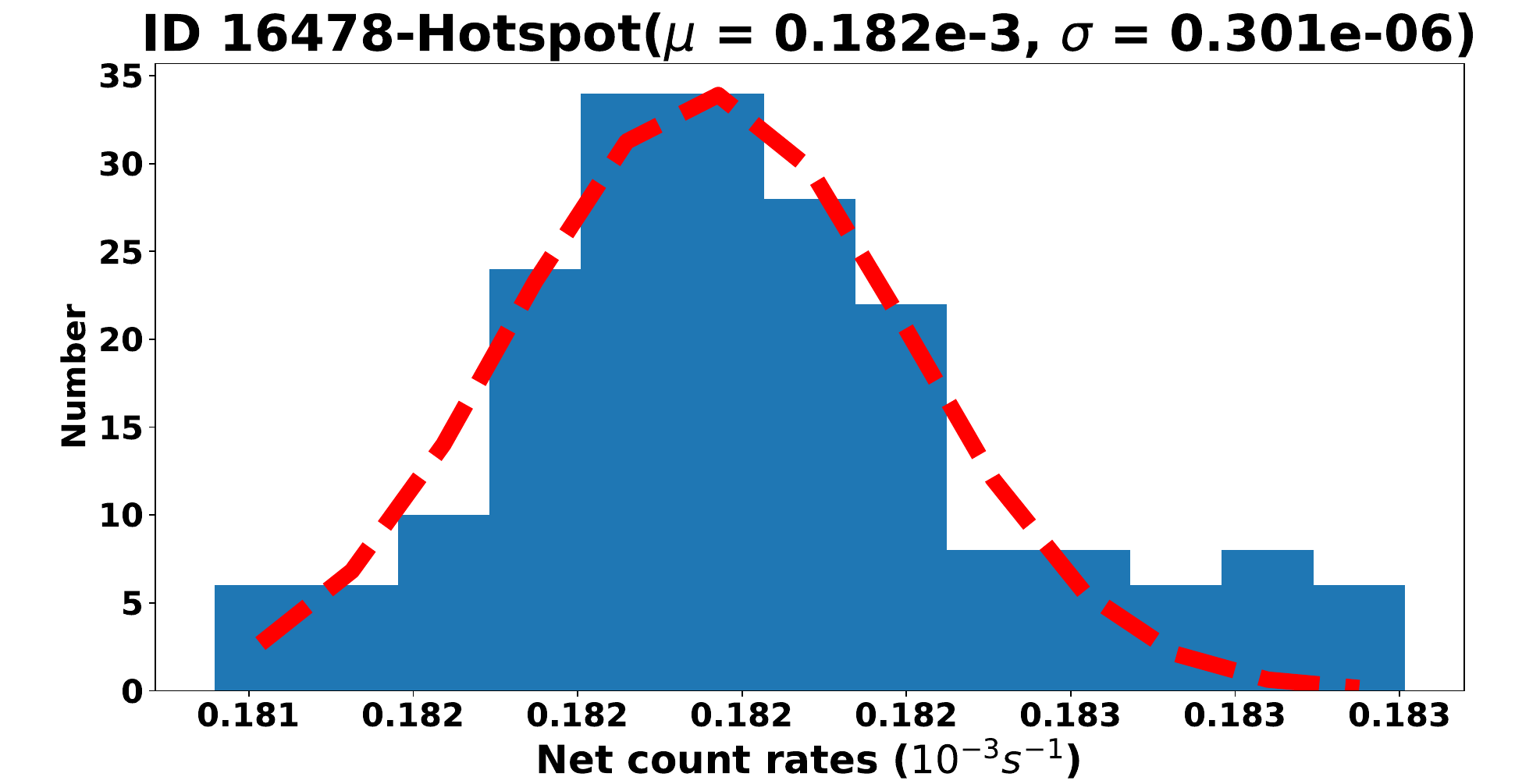}
\includegraphics[width=0.24\textwidth]{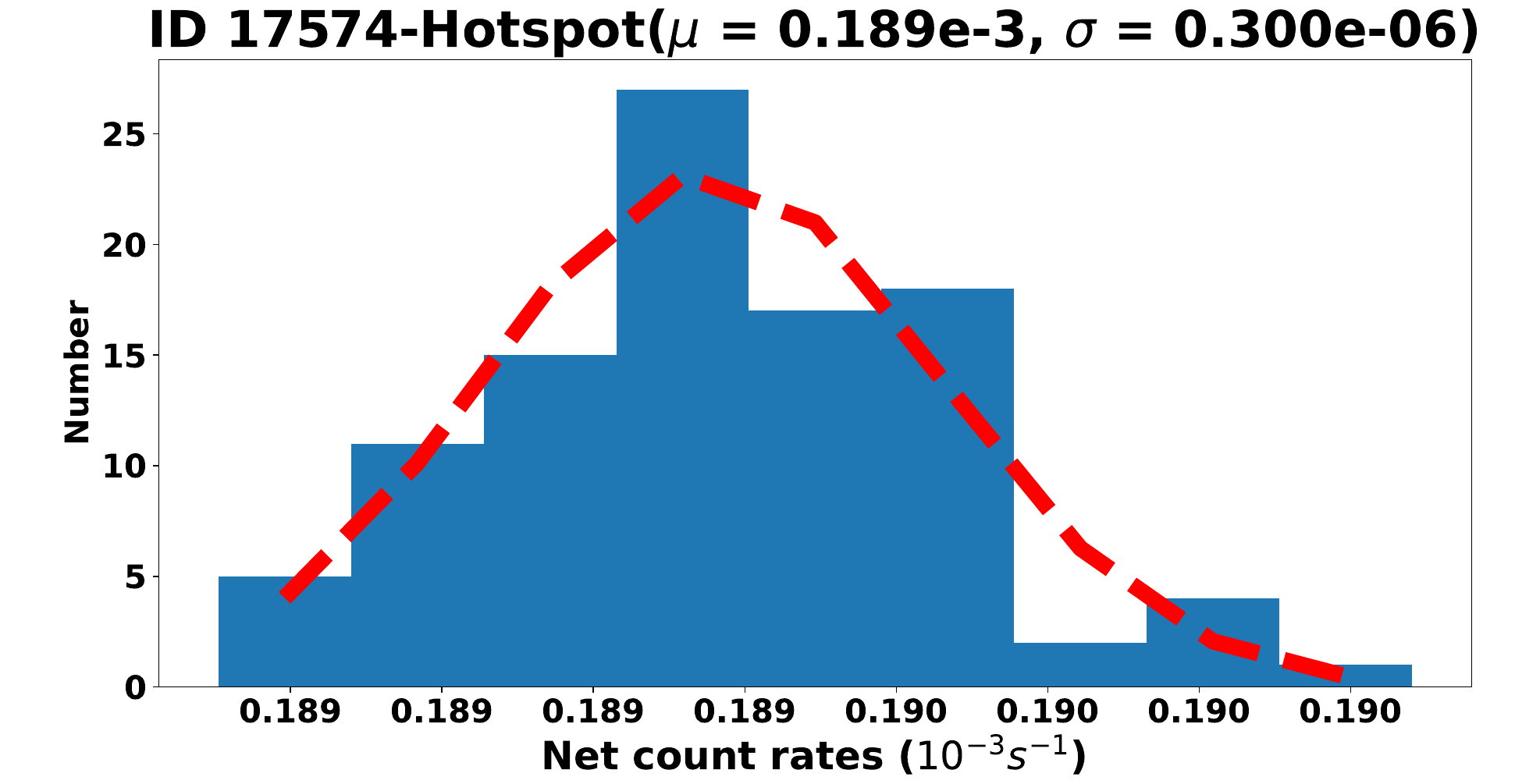}
\caption{Histograms of the net count rates calculated for the HS region on the deconvolved exposure-corrected {\it Chandra} images of the W hotspot in Pictor\,A, at 1\,px resolution, for all the analyzed ObsIDs. Each panel corresponds to 100 random realizations of the PSF for a given ObsID.}
\label{fig:histogram-all}
\end{figure*}

\begin{figure*}[h]
\centering 
\centering 
\includegraphics[width=0.24\textwidth]{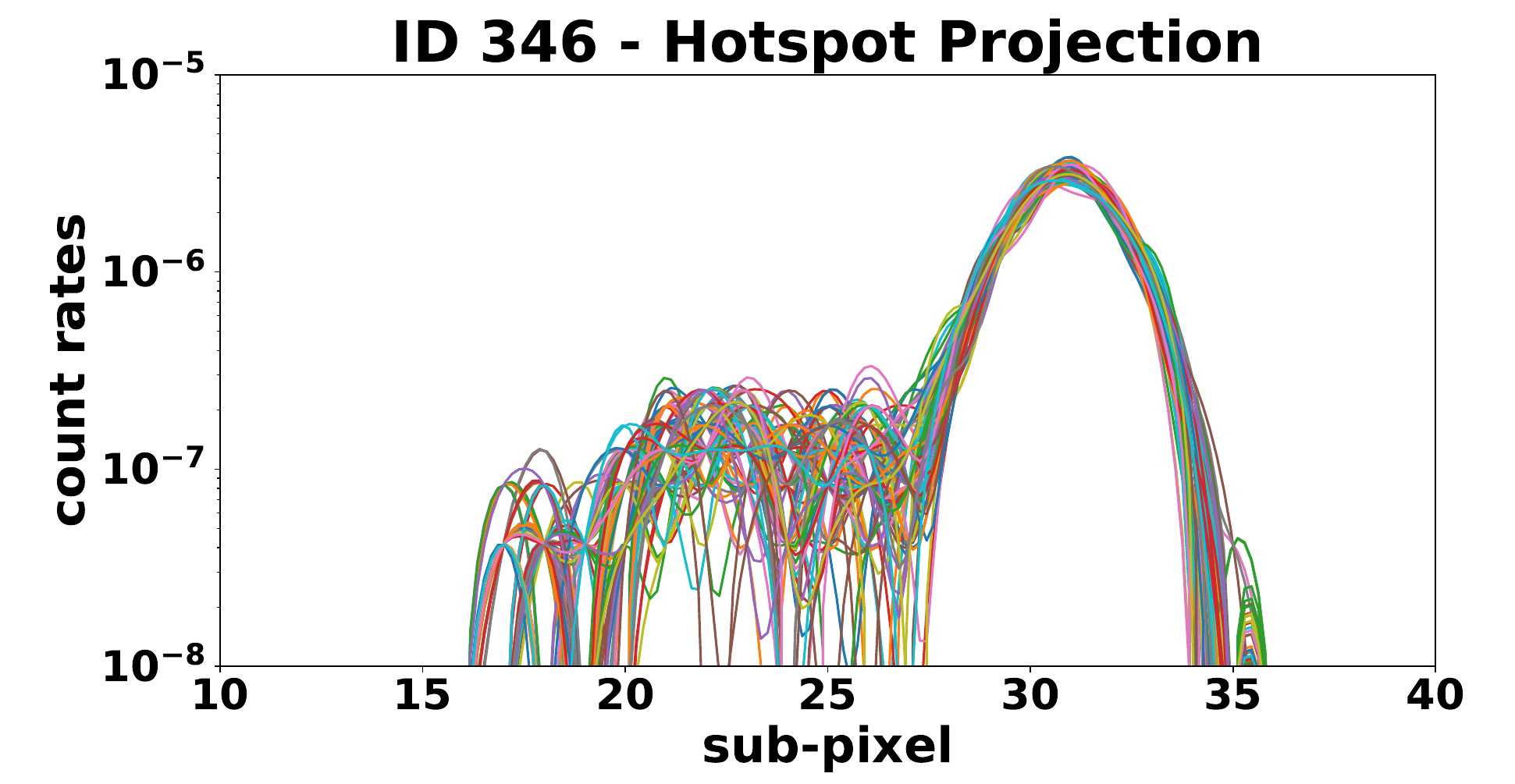}
\includegraphics[width=0.24\textwidth]{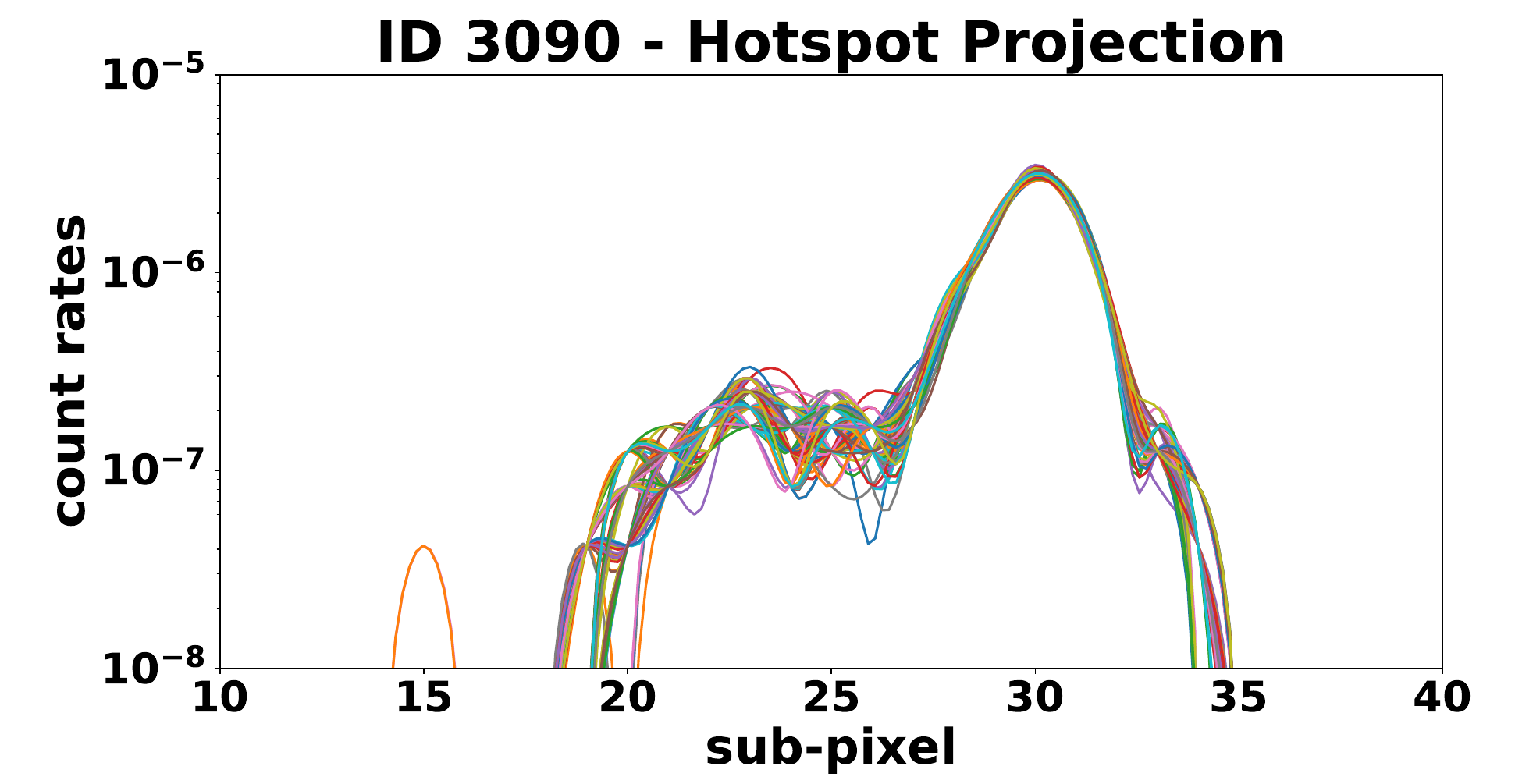}
\includegraphics[width=0.24\textwidth]{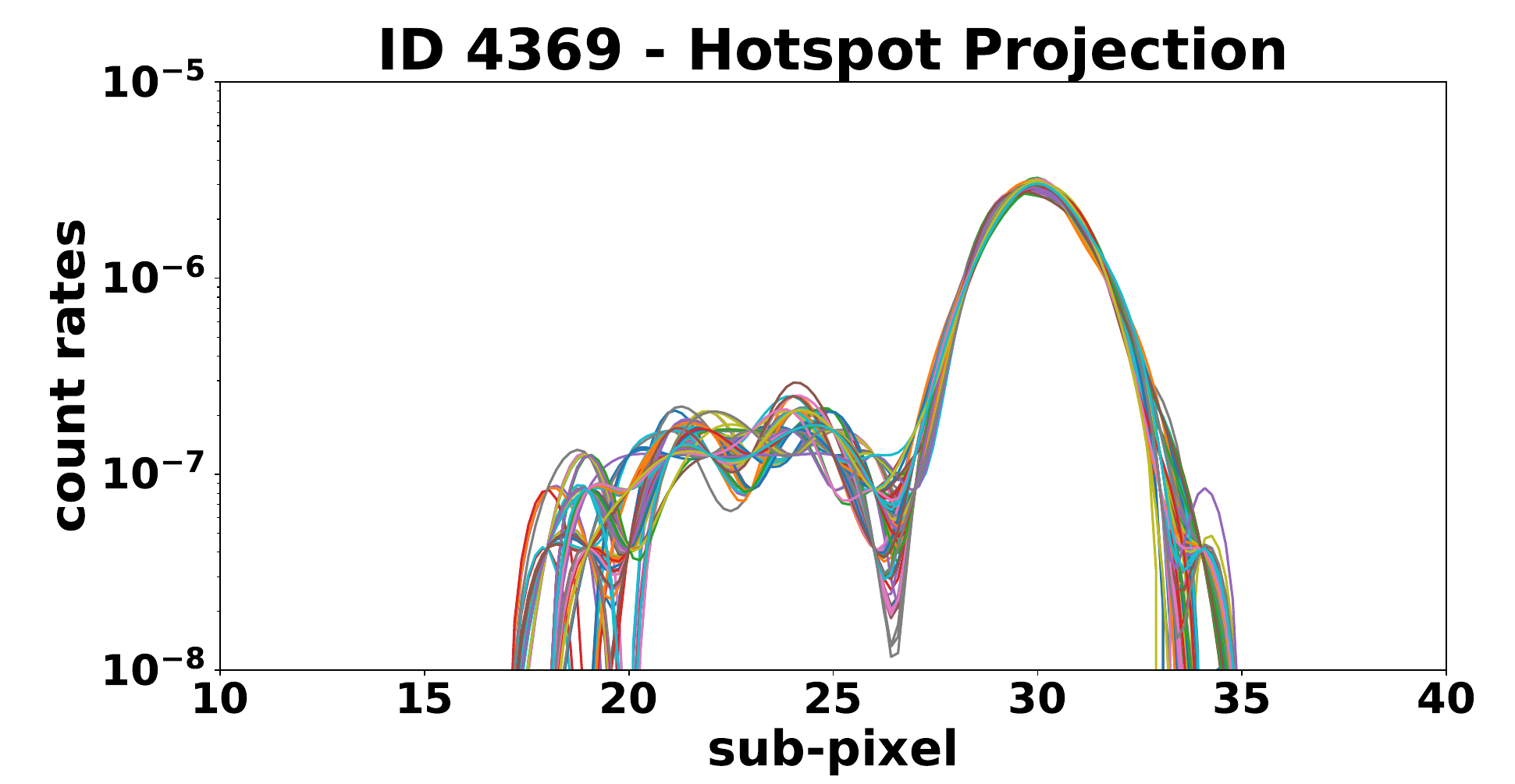}
\includegraphics[width=0.24\textwidth]{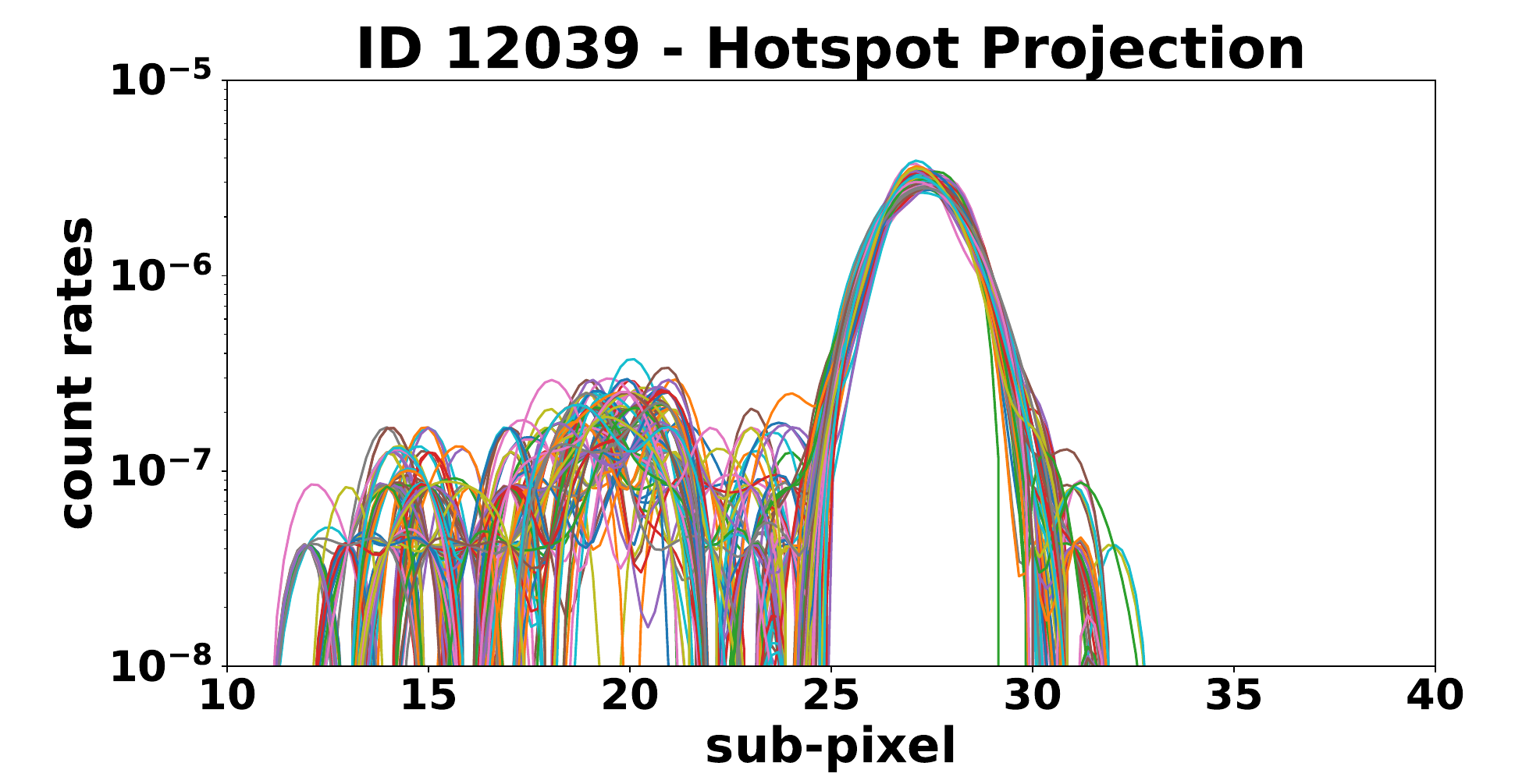}
\includegraphics[width=0.24\textwidth]{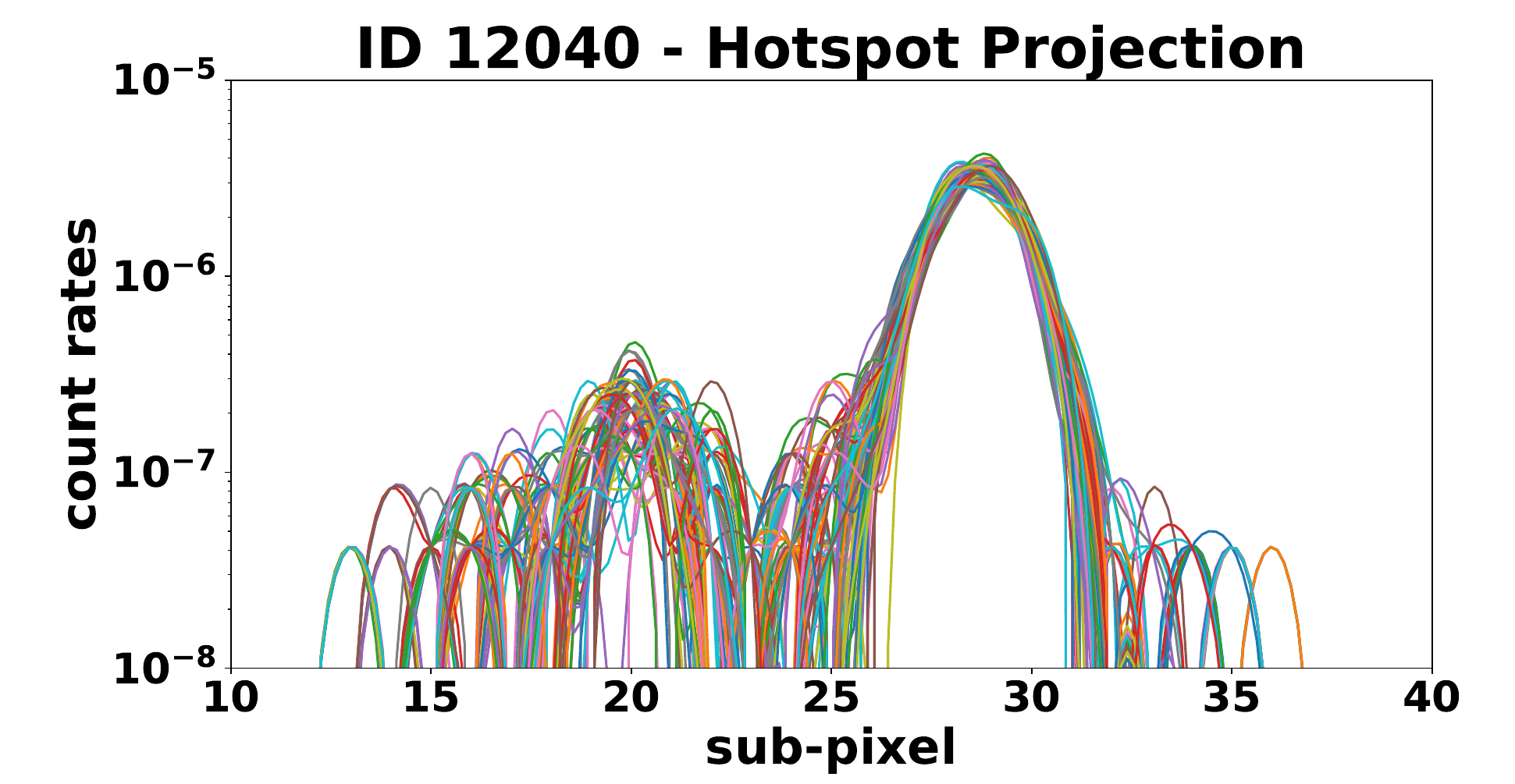}
\includegraphics[width=0.24\textwidth]{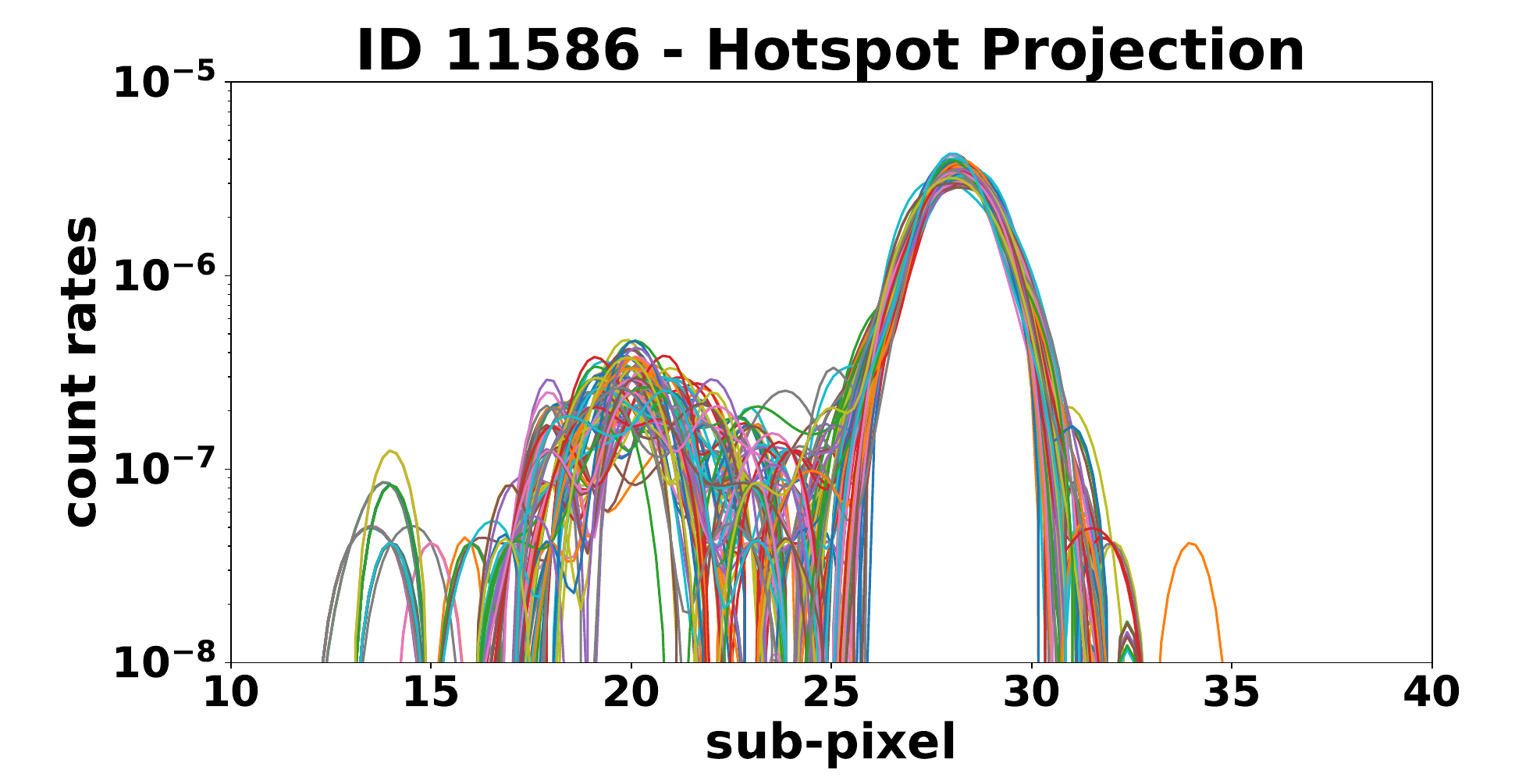}
\includegraphics[width=0.24\textwidth]{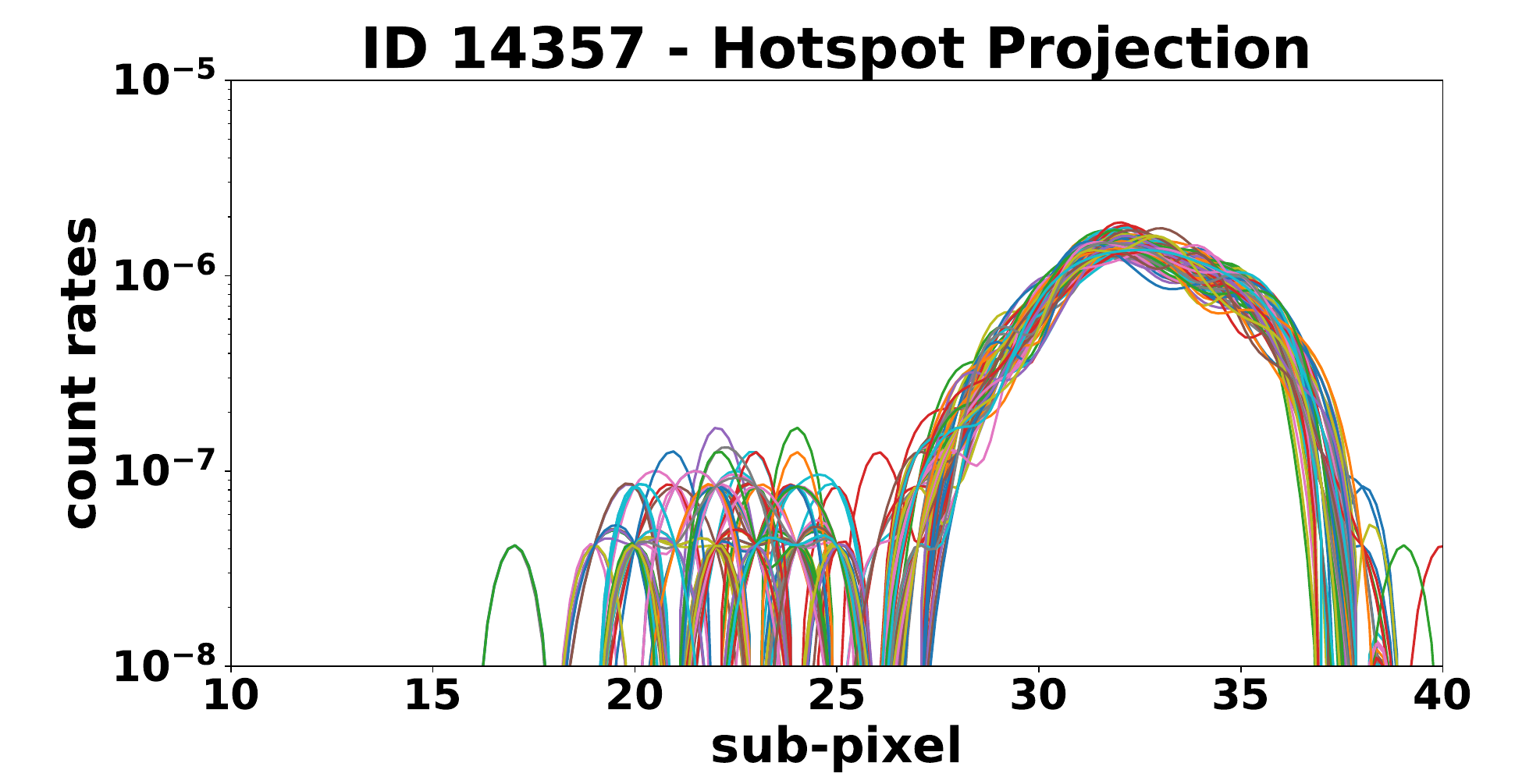}
\includegraphics[width=0.24\textwidth]{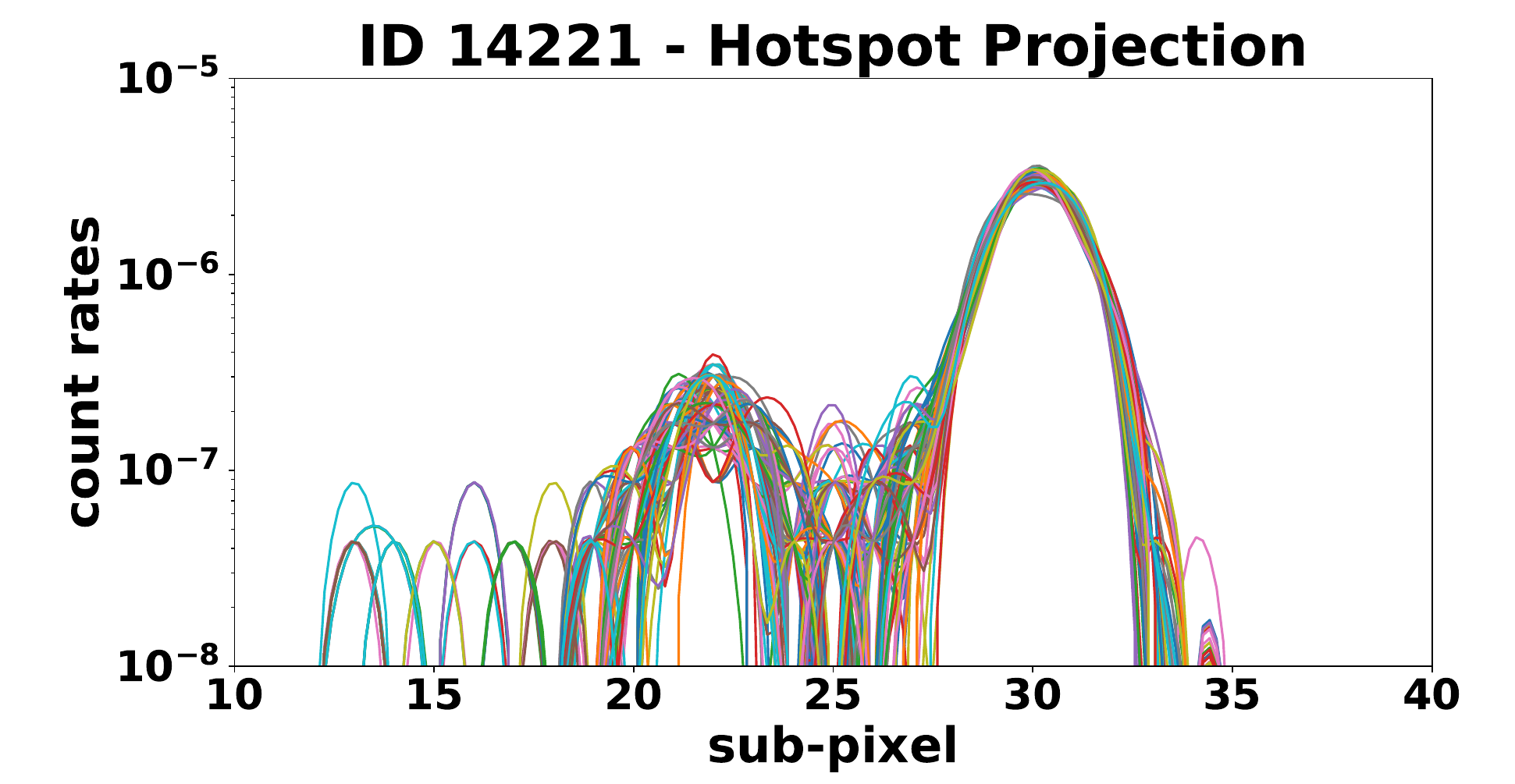}
\includegraphics[width=0.24\textwidth]{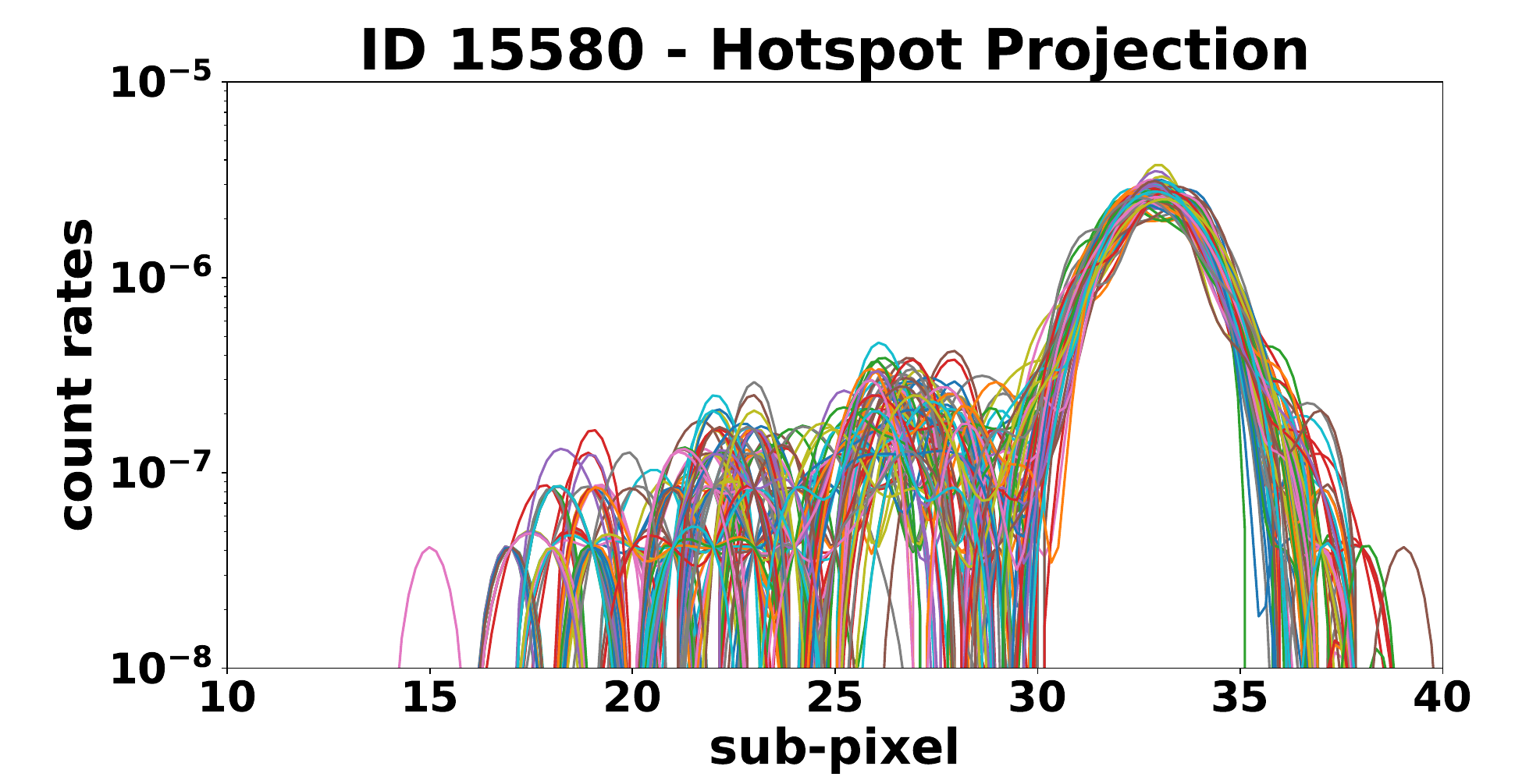}
\includegraphics[width=0.24\textwidth]{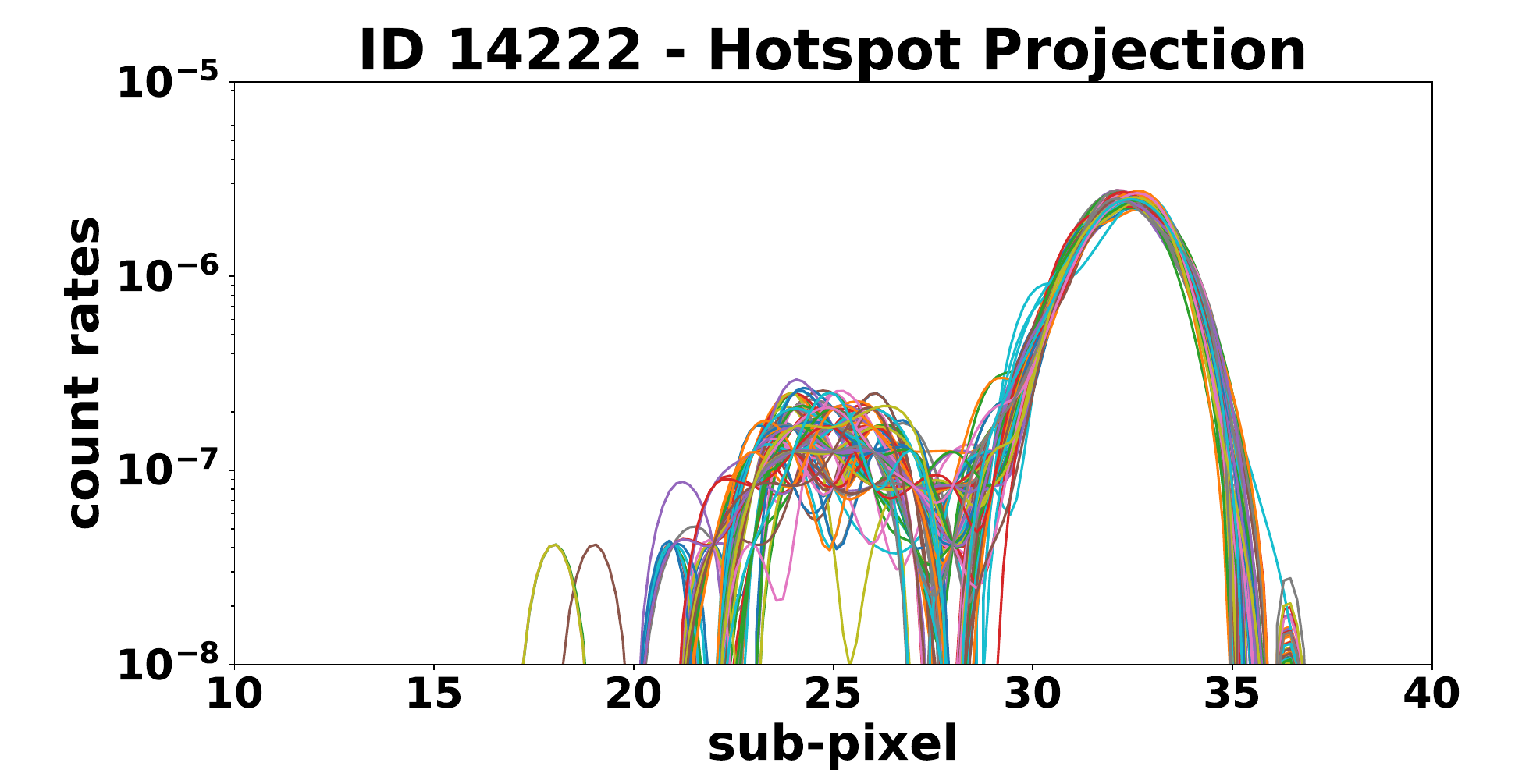}
\includegraphics[width=0.24\textwidth]{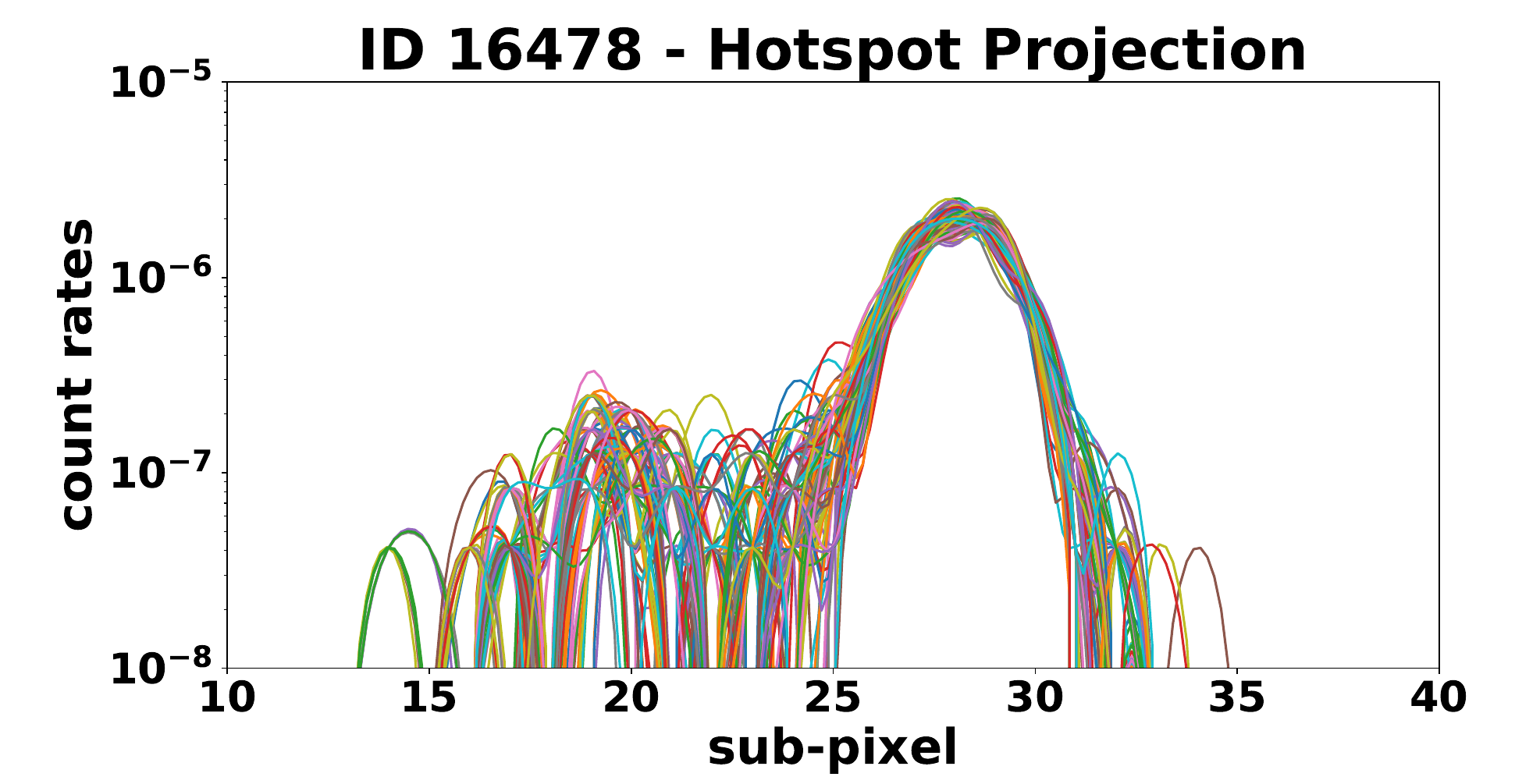}
\includegraphics[width=0.24\textwidth]{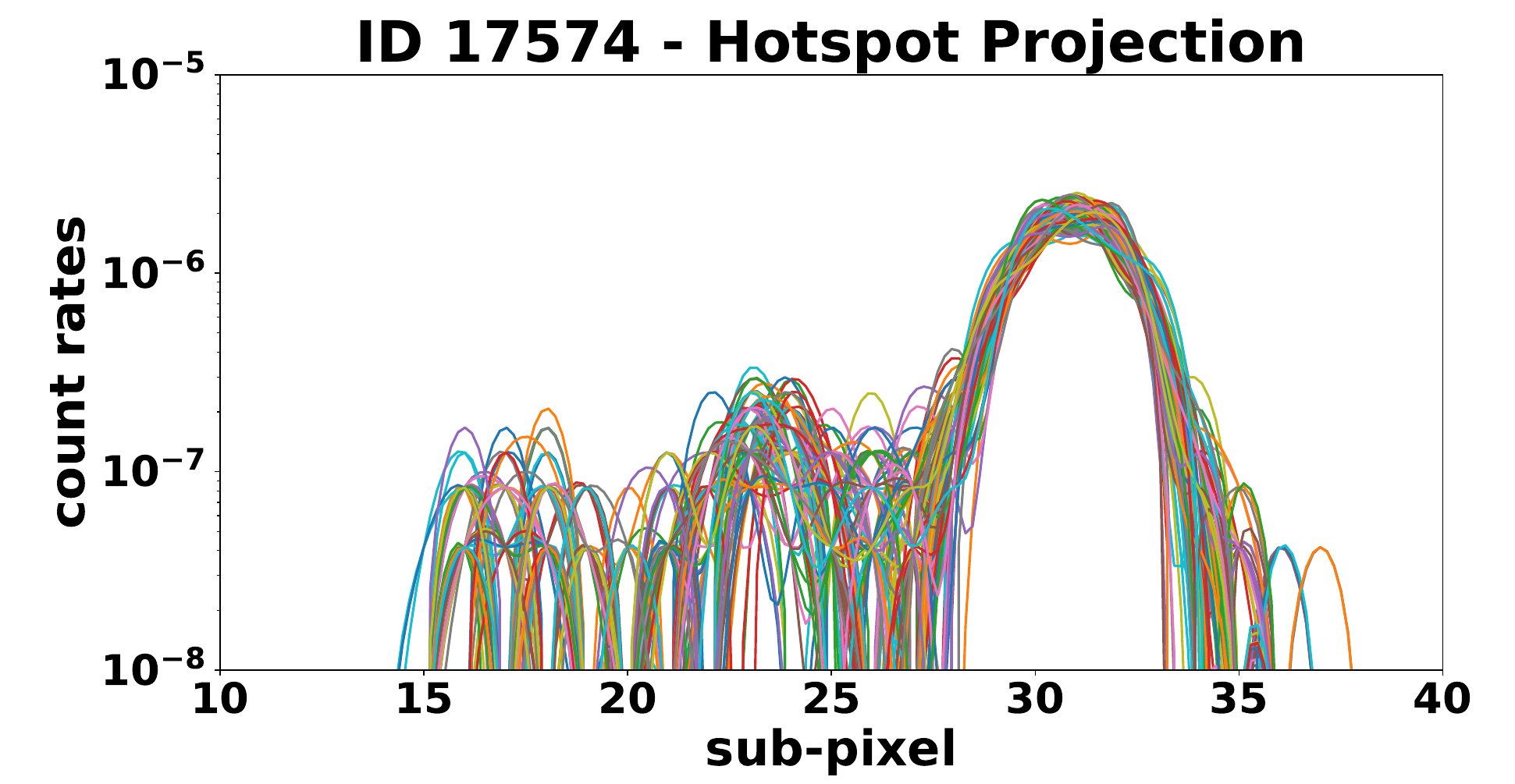}
\caption{Integrated intensity profiles along the major axis of the N region displayed on Figure\,\ref{fig:regions}, for all the analyzed ObsIDs, at 0.5\,px resolution. Each curve on the panels corresponds to one single realization of the PSF for a given ObsID.}
\label{fig:profiles-all-subpx}
\end{figure*}

\end{document}